\documentclass[twocolumn,aps,prb,nofootinbib,superscriptaddress,floatfix,hyperref=pdftex]{revtex4-1}

\usepackage[colorlinks,bookmarks=false,citecolor=blue,linkcolor=blue,urlcolor=blue,hyperfootnotes=false]{hyperref}

\usepackage[utf8]{inputenc} 

\usepackage[usenames]{xcolor}

\usepackage[T1]{fontenc}

\usepackage{amsmath,amssymb,amsfonts,accents}

\usepackage{mathtools,mathdots,mathpple}

\usepackage{braket,mathrsfs,bbold,bm}

\usepackage{eucal,bigints,dsfont}

\usepackage{times,color,lipsum}

\usepackage{tikz,graphicx,dcolumn}

\usepackage{wasysym,relsize}

\setcitestyle{numbers,square}

\DeclareMathSizes{10}{9.3}{7}{5}

\hypersetup{colorlinks,citecolor=blue,filecolor=black,linkcolor=red,urlcolor=blue}

\makeatletter

\newcommand{\beq}{\begin{equation}}
\newcommand{\eeq}{\end{equation}}

\newcommand{\bqa}{\begin{eqnarray}}
\newcommand{\eqa}{\end{eqnarray}}

\newsavebox{\mybox}

\global\long\def\ket#1{|#1\rangle}

\newcommand{\smath}[1]{\mbox{\scriptsize\ensuremath{#1}}}

\newcommand{\tmath}[1]{\mbox{\tiny\ensuremath{#1}}}

\newcommand{\tsp}[0]{\hspace{0.7pt}}

\newcommand{\sprm}[0]{\tsp\prime}

\newcommand{\bsm}[1]{\boldsymbol{#1}}

\newcommand{\phdg}[0]{\vphantom{\dagger}}

\makeatother

\begin{document}

\title{Strange metallicity in an antiferromagnetic quantum critical model:\\
A sign-problem-free quantum Monte-Carlo study}

\author{Rafael M. P. Teixeira}
\affiliation{Instituto de Física, Universidade Federal de Goiás,
    74.001-970, Goiânia-GO, Brazil}

\author{Catherine Pépin}
\affiliation{Institut de Physique Théorique, Université Paris-Saclay,
    CEA, CNRS, F-91191 Gif-sur-Yvette, France}

\author{Hermann Freire}
\affiliation{Instituto de Física, Universidade Federal de Goiás,
    74.001-970, Goiânia-GO, Brazil}
  
\date{\today}

\begin{abstract} 

We compute transport and thermodynamic properties of a two-band spin-fermion model describing itinerant fermions in two dimensions interacting via $Z_2$ antiferromagnetic quantum critical fluctuations by means of a sign-problem-free quantum Monte-Carlo approach. We show that the phase diagram of this model indeed contains a $d$-wave superconducting phase at low enough temperatures. However, a crucial question that arises is whether a non-Fermi-liquid metallic regime exists above $T_c$ exhibiting hallmark strange-metal transport phenomenology. Interestingly, we find that this version of the model describes a non-Fermi-liquid metallic regime that displays an approximately $T$-linear resistivity above $T_c$ for a strong fermion-boson interaction. Using Nernst-Einstein relation, our QMC results also show that this strange metal phase exhibits a crossover from being characterized by a charge compressibility given approximately by $\chi_{\tsp c}\sim 1/T$ at high temperatures to being described by a charge diffusivity consistent with the scaling $D_{c}\sim 1/T$ at low temperatures. Therefore, our work adds support to the view that the $Z_2$ antiferromagnetic spin-fermion model at strong coupling can be considered a minimal model that describes both unconventional superconductivity and strange metallicity, which are fundamentally interconnected in many important strongly-correlated quantum materials.

\end{abstract}

\maketitle

\section{Introduction}

Quantum criticality is a common theme in many strongly correlated quantum materials\cite{Sachdev-CUP(2011)} such as, e.g., the high-$T_c$ cuprates \cite{Hinkov-S(2008),Taillefer-N(2010),Ando-PRL(2002),Matsuda-NP(2017),Kaminski-N(2002),Bourges-CRP(2016)}, heavy-fermion compounds \cite{Thompson-S(2008)} and iron-based superconducting compounds \cite{Kivelson-PRB(2008),Davis-S(2010),Fisher-S(2012)} (to name only a few systems). This is due to the fact that the unconventional superconducting phases exhibited by those systems always occur close to either one or more symmetry-broken phases in their corresponding phase diagrams, whose critical order-parameter fluctuations are believed to provide the underlying mechanism that mediates the pairing between the fermions \cite{RPhys2012}. However, despite this great potential for universal classification in very different materials, quantum critical models are famously difficult to be solved analytically, even in simplified large-$N$ flavor generalizations of such systems. This owes to the fact that the interactions of these models are relevant parameters in the renormalization group sense and typically flow to strong coupling at low energies \cite{Abanov_Chubukov,Metlitski-PRBb(2010)}. Therefore, in general, perturbative approaches to calculate their physical properties at relevant temperature scales cannot be used in a reliable manner. As a consequence, non-perturbative approaches have become of paramount importance for describing these models in recent years. 

One celebrated quantum critical model is the spin-fermion model in two spatial dimensions \cite{Abanov_Chubukov}. It considers itinerant fermions in the vicinity of a Fermi surface (FS) interacting via antiferromagnetic (AFM) fluctuations that effectively carry momentum close to $({\pi,\pi})$. Recently, it has been investigated analytically by several authors with many important results. In this regard, we point out the work in Ref. \cite{Metlitski-PRBb(2010)} who implemented a renormalization group analysis combined with a $1/N$-expansion (with $N$ being the number of fermionic flavors) for this model. As a result, they found that although the $1/N$-expansion ultimately fails for this problem due to the emergence of strong quantum fluctuations at low energies, they obtained interesting renormalizations of several parameters of the model. As an example, it was demonstrated that the Fermi-liquid theory breaks down near the so-called hot-spots \cite{Metlitski-PRBb(2010)}, which refer to special points of the model in momentum space that represent the intersection of the underlying FS with the antiferromagnetic zone boundary. At weak coupling, the hot-spots are conjectured to effectively control the universal properties of the spin-fermion model, i.e., different models will belong to the same universality class in the low-energy limit provided that the angles between the Fermi velocities at the hot-spots are the same.

Later on, in Ref. \cite{SSLee_Exact} a self-consistent non-perturbative analytical strategy was proposed, building on previous results by the same authors \cite{Lunts_Lee}, to solve the spin-fermion model with $O(3)$ symmetry near the hot-spots using an emergent control parameter: the degree of local nesting at those points. As a result, they obtained a strong-coupling infrared fixed point at very low energies in the model, which is associated with: {$i)$} a bosonic dynamical critical exponent given by $z = 1$, {$ii)$} a consequent  emergent nesting at the hot-spots and {$iii)$} the existence of a finite bosonic anomalous dimension in the theory. 

In contrast, we will focus here on the spin-fermion model with $Z_2$ symmetry. One of the motivations for the present study is that this model is expected to have a lower superconducting transition temperature compared to the same model with $O(3)$ symmetry. This will give us a large temperature window in which we will be able to characterize the normal state of this model. In this regard, there is an important discussion in the literature (see, e.g., Refs. \cite{Metlitski_2015,Wang_Chubukov,Borges_Lee}) about whether the formation of a superconducting phase in quantum critical models preempts the emergence of non-Fermi liquid features at low temperatures or if a non-Fermi liquid is capable of surviving within a sizable temperature window above $T_c$. Although there was recently great progress on this question regarding the antiferromagnetic spin-fermion with $O(3)$ symmetry \cite{SSLee_ARCMP,Borges_Lee}, the same study for $Z_2$ spins has not yet been carefully investigated to our best knowledge.

Another important non-perturbative approach to this problem is provided by unbiased numerical simulations such as, e.g., the sign-problem-free quantum Monte-Carlo (QMC) method. This line of research was initiated in recent years in Refs. \cite{Sci2012,Schattner:2016dw,Berg_2019,Berg_cond-mat,ZYMeng} and it has now been established, e.g., that a two-band version of the spin-fermion model describes a high-$T_c$ superconducting phase \cite{Schattner:2016dw,Berg_cond-mat} with a pairing gap consistent with $d$-wave symmetry (similar in this respect to the cuprate superconductors). The choice of an effective two-band model that preserves the structure of the hot-spots is instrumental, since it was demonstrated that there exists an antiunitary operator in this system that renders the numerical QMC simulations fermionic sign-problem-free \cite{Sci2012}.

To further elucidate the physical mechanism underlying the formation of the superconducting state, another recent QMC work on this model was given in Ref. \cite{Wang:2016tr}, in which a comparison between the numerical QMC method and the field-theoretical perturbative Eliashberg approximation was made. As a result, those authors have demonstrated numerically that from weak to intermediate couplings (compared to the non-interacting bandwidth of the model), the hot-spot-only Eliashberg approximation to the problem gives surprisingly good results concerning, e.g., the critical temperature of the corresponding superconducting phase \cite{Wang:2016tr}. Despite this, at very strong couplings, this comparison starts to become worse, thus showing that the perturbative Eliashberg approximation eventually breaks down for large enough couplings in the spin-fermion model. 

Transport properties are of course also of crucial interest in this context, since those quantities provide another important way to characterize the unconventional phases that emerge in these systems. In this respect, we note that transport theories for AFM quantum criticality now have a long history in the literature (see, e.g., Refs. \cite{Hlubina,RoschPRL,Patel_Strack,Hartnoll_Hofman,Schmalian,Maslov_Chubukov_2014}). This problem was addressed by many authors using different analytical methods such as, e.g., the Boltzmann equation method \cite{Hlubina,RoschPRL,Patel_Strack} and the Kubo formula \cite{Hartnoll_Hofman,Schmalian,Maslov_Chubukov_2014}. From a weak-coupling perspective, Ref. \cite{Hlubina} showed that in the clean limit, since only the hot spots at the underlying FS couple efficiently to the AFM fluctuations, the remaining regions of the FS would essentially remain cold. This would lead to a conventional Fermi liquid transport due to the short-circuiting of the unconventional contribution to the resistivity originated from the hot-spots. Later, in Ref. \cite{RoschPRL}, a non-Fermi-liquid transport result was obtained in the model by introducing additionally weak disorder, which effectively changes the balance of hot-spot and cold-region contributions in the system.

However, from a strong-coupling viewpoint, another scenario has recently been put forward, starting from other transport theories\cite{Hartnoll-PRB_2013,Patel-PRB,Lucas-PRB,Hartnoll_PRB_2014,Freire-AP_2020,Freire-AP_2017,Freire-EPL,Freire-EPL_2018,Freire-AP_2014,Zaanen-CUP,Sachdev-MIT,freire_2020,ips-hermann2,Freire_Mandal_2022,Berg2019,Berg2022,Chowdhury_RMP} that draw inspiration from non-perturbative calculations in holographic models of metallic states (see, e.g., Refs. \cite{Zaanen-CUP,Sachdev-MIT,Chowdhury_RMP}). This new perspective is based on the memory-matrix approach\cite{Forster-HFBSCF(1975),Goetze_Woefle,Rosch_Andrei,Chowdhury_RMP} that successfully captures the hydrodynamic regime, which is expected to describe the non-equilibrium dynamics of the strange metal phase. In this point of view, due to the strong coupling nature of the spin-fermion interaction in two spatial dimensions \cite{Metlitski-PRBb(2010)}, the bosonic order parameter fluctuations will not only couple to the hot-spots, but it can also couple effectively to the remaining parts of the underlying FS via composite operators \cite{Vicari_2008,Hartnoll_Hofman}. Consequently, the whole FS is expected to become ``lukewarm'', which could then lead in some cases to non-Fermi-liquid behavior in the transport coefficients.

In the present paper, we investigate transport and thermodynamic properties of the $Z_2$ AFM spin-fermion model using the sign-problem-free QMC method. The main aim of our paper is to perform nonperturbative QMC simulations on this model for stronger couplings in a regime where the Eliashberg approximation, in principle, is not expected to yield qualitatively good results. Specifically, we will focus on describing the metallic state that exists above $T_c$ in the corresponding phase diagram. In this way, our goal here will be to address the following fundamental questions regarding the present problem: (i) Can a superconducting phase with $d$-wave symmetry exist in the $Z_2$ AFM spin-fermion model at low enough temperatures? (ii) Can a strange metal with
$T$-linear resistivity emerge in the model above the $d$-wave superconducting phase? (iii) What is the mechanism that drives the formation of this non-Fermi liquid metallic state?

Therefore, the remainder of this work will be organized as follows: In Sec. \ref{sec:II}, we will define the $Z_2$ AFM spin-fermion model that we want to investigate. In Sec.  \ref{sec:III}, we briefly explain the sign-problem-free QMC methodology applied to this model. Next, in Sec.  \ref{sec:IV}, we will present our numerical results regarding this investigation. In Sec. \ref{sec:V}, we end with the summary and an outlook of our present study. Lastly, in the Appendix we provide more details about the finite size effects on our simulation results. 

\section{Lattice model}\label{sec:II}

We will consider an effective two-band (or two-flavor) spin-fermion model with fermions from each band assigned with a band/flavor index $\alpha=1,2$ where the interaction between $\alpha\tsp$-fermions and $\alpha^{\sprm}$-fermions emerge from their coupling with a $Z_2$ AFM order parameter field. The Euclidean action of this system is written as a sum of two contributions: $S[\bar{\psi},\psi,\varphi]=S_{\psi}[\bar{\psi},\psi,\varphi]+S_{\varphi}[\varphi]\tsp$. In this manner, the partition function is given by the following coherent-state path-integral:
\begin{align}
    Z & =\!\int\!D(\bar{\psi},\psi,\varphi)\;
    \mbox{e}^{\,-S[\bar{\psi},\psi,\varphi]/\tsp\hbar}\nonumber\\[0.9mm]
    & =\!\int\!D\varphi\left\{\mbox{e}^{\,-S_{\varphi}[\varphi]/\tsp\hbar}\!
    \int\!D(\bar{\psi},\psi)\;\mbox{e}^{\,-S_{\psi}
    [\bar{\psi},\psi,\varphi]/\tsp\hbar}\right\}
    \nonumber\\[-0.6mm]
    & =\!\int\!D\varphi\;\mbox{e}^{\,-S_{\varphi}[\varphi]/\tsp\hbar}\;
    \textrm{\large Tr}_{\tsp\psi}\!\left[\,
    \lim_{\Delta\tau\rightarrow\, 0^{\vphantom{X}}}
    \;\prod_{m=1^{\vphantom{X}}}^{M}
    e^{-\Delta\tau\, H(\tau_{m})/\tsp\hbar}\,\right],\label{Lagrangian}
\end{align}  
where $\tau_{m}=m\tsp\Delta\tau$ (with $m=1,2,\dots,M$ and $\Delta\tau=\hbar\beta/M\tsp$) represent discrete values of the imaginary-time $\tau\in[0,\hbar\beta]$, in which $\beta=1/k_{B}T$ is the inverse temperature (for simplicity, we set $\hbar=1$ and $k_{B}=1$ from now on). In this formalism, the Grassmann variables ($\bar{\psi},\psi$) and the bosonic field ($\varphi$) are $\tau$-\tsp dependent. In terms of fermionic creation (annihilation) operators $c^{\tsp\tmath{\dagger}}_{\alpha,\tsp i,\tsp s}$ ($\,c^{\tmath{\phdg}}_{\alpha,\tsp i,\tsp s}\,$) corresponding to the Grassmann variables $\bar{\psi}_{\alpha,\tsp i,\tsp s}$ ($\,\psi_{\alpha,\tsp i,\tsp s}\,$), the $\tau$-\tsp dependent Hamiltonian $H(\tau)$ reads:
\begin{align}
    H( & \tau)=-\sum_{\alpha}\sum_{i,j}\sum_{s}\left[\tsp
    t^{(\alpha)}_{\,ij}+\delta_{ij}\,\mu\tsp\right]
    c^{\dagger}_{\alpha,\tsp i,\tsp s}\,c^{\phdg}_{\alpha,j,\tsp s}
    \label{FHam}\\[-0.5mm]
    & +\lambda\,\sum_{i}\,e^{\,\text{i}\hspace{0.5pt}\bsm{Q}\,\cdot\,\bsm{r}_{i}}\,
    \varphi_{i}(\tau)\!\left[\,c^{\dagger}_{1,\tsp i,\tsp\uparrow}\,
    c^{\phdg}_{2,\tsp i,\tsp\downarrow}\!+c^{\dagger}_{2,\tsp i,\tsp\uparrow}\,
    c^{\phdg}_{1,\tsp i,\tsp\downarrow}\tsp\right]+\textrm{H.c.},\nonumber
\end{align}
where $\alpha=1,2$ are the band indices, $s=\uparrow,\downarrow$ are the spin indices, $\bsm{r}_{i}$ (for $i=1,2,\dots,N_{s}$) is the $i$-th site position on a two-dimensional (2D) square lattice of $N_{s}=L\times L$ sites ($\,j$ is defined in the same way) with spacing $a\tsp$, $t^{\tsp\tmath{(\alpha)}}_{\,\tmath{ij}}$ are the hopping parameters associated with the $\alpha\tsp$-$\tsp$band, $\mu$ is the chemical potential, $\lambda$ is the Yukawa coupling parameter, and $\bsm{Q}=(\pi/a,\pi/a)$ is the wavevector associated with the commensurate SDW order whose fluctuations in the lattice are represented by $\varphi_{i}\tsp$. The action can be written as an imaginary-time integral of the fermionic and bosonic parts of the Lagrangian of the system: $S[\bar{\psi},\psi,\varphi]=\int_{\smath{0}^{\vphantom{X}}}^{\tsp\smath{\beta}}d\tau \left[\,L_{\tsp\psi}(\bar{\psi},\psi,\varphi\tsp;\tau)+L_{\tsp\varphi}(\varphi\tsp;\tau)\,\right]\tsp$. The term $L_{\tsp\psi}$ is given by
\vspace{-1mm}
\begin{equation}
    L_{\tsp\psi}(\bar{\psi},\psi,\varphi\tsp;\tau)=
    \sum_{\alpha,\tsp i,\tsp s}\!\bar{\psi}^{\phdg}_{\alpha,\tsp i,\tsp s}(\tau)\,
    \partial_{\tau}\tsp\psi^{\phdg}_{\alpha,\tsp i,\tsp s}(\tau)+
    \mathcal{H}(\bar{\psi},\psi,\varphi\,;\tau),
    \label{FLag}
\end{equation}
with $\mathcal{H}(\bar{\psi},\psi,\varphi\,;\tau)$ being the coherent-state path-integral form of the Hamiltonian in Eq. \eqref{FHam}. The bosonic part $L_{\tsp\varphi}$ has the following Ginzburg-Landau (GL) form ($\varphi_{i}$ depends on $\tau$):
\begin{align}
    L_{\tsp\varphi}(\varphi\tsp;\tau)
    & =\dfrac{1}{2}\sum_{i\tsp=\tsp1^{\vphantom{X}}}^{N_{s\vphantom{x_{x}}}}
    \dfrac{1}{c^{\,2}}\!\left(\dfrac{d\varphi_{i}}{d\tau}\right)^{\!\!2}\!+
    \dfrac{1}{2}\!\sum_{\left\langle i,j\right\rangle^{\vphantom{X}}}\!\left(
    \varphi_{i}-\varphi_{j}\right)^{2}\nonumber\\[-1.5mm]
    & +\sum_{i\tsp=\tsp1^{\vphantom{X}}}^{N_{s\vphantom{x_{x}}}}
    \left(\dfrac{r}{2}\tsp\varphi_{i}^{\,2}+
    \dfrac{u}{4}\tsp\varphi_{i}^{\,4}\right),
    \label{BLag}
\end{align}
with $r$ being a parameter (which can be related to either doping or to an applied external pressure) that tunes the system through a SDW quantum critical point, $c$ is the bare bosonic (SDW) velocity, and $u$ is the quartic coupling. The GL action $S_{\varphi}[\varphi]=\int_{\smath{0}^{\vphantom{X}}}^{\tsp\smath{\beta}}d\tau\,L_{\tsp\varphi}(\varphi)$ can be considered to be the result of the process of integrating out high-energy electronic degrees of freedom, in which the time and amplitude fluctuations of the bosonic field are assumed to be small so that the first-order time and spatial derivatives are enough for the effective description of the field near the quantum critical point (moreover, the overall amplitude of the field is also assumed to be small to ensure that the leading-order approximation of the GL theory is valid).

\begin{figure}[t]
    \centering
    \includegraphics[width=0.8\columnwidth]{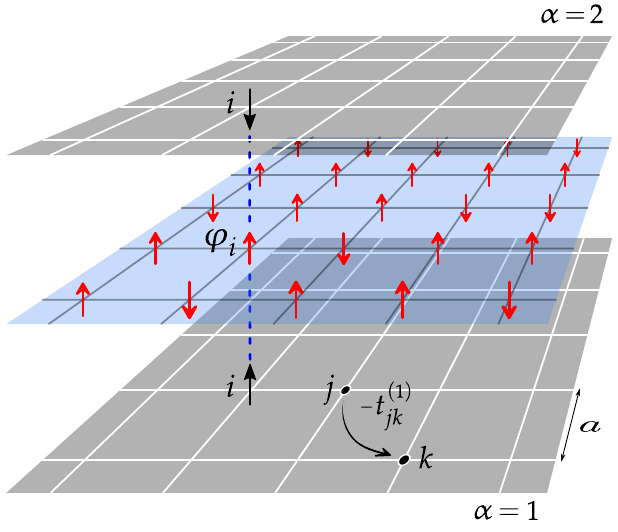}
    \caption{Schematic representation of the two-band spin-fermion model with a $Z_2$ AFM order parameter field. Each $\alpha\tsp$-$\tsp$band ($\alpha=1,2$) is defined by the bare dispersion $\varepsilon_{\bsm{k}}^{\tsp\tmath{(\alpha)}}$ associated with an independent layer representing a 2D lattice system ($a$ is the lattice spacing), where the hopping parameters of the fermions with flavor $\alpha$ are given by  $t^{\tsp\tmath{(\alpha)}}_{\,\tmath{ij}}$. The interaction (illustrated by a dashed line) between the fermions of different flavors with opposite spins at the $i$-th site in both layers occurs via an intermediate layer composed by SDW fluctuations represented by the order parameter field $\varphi_{i}$ .}
    \label{Fig0}
\end{figure}

The present model can be represented as a system made of two parallel layers labeled by the flavor index $\alpha$. With no interaction (i.e., $\lambda=0$), these layers consist of two independent lattices where the fermions hop around and the associated bare energy dispersion $\varepsilon_{\tmath{\bsm{k}}}^{\tmath{(\alpha)}}$ defines the $\alpha$-band that depends on the choice of $t^{\tmath{(\alpha)}}_{\,\tmath{ij}}$ parameters. In this picture, the two-band system is formed when these layers are coupled by the Yukawa interaction ($\lambda>0$) so that fermions of different flavors with opposite spins at the $i$-th site in both layers interact via the SDW fluctuations. This interaction is depicted in Fig. \ref{Fig0} (illustrated by a dashed vertical line) where the $Z_2$ fluctuations are represented schematically by the $\varphi_{i}$ field in an intermediate layer between the $\alpha=1$ and $\alpha=2$ layers. In the context of determinantal QMC simulations, this two-band model with an AFM order parameter field turns out to be sign-problem-free, i.e., the fermionic determinant is always positive definite (we mention here that the single-band version of this model suffers from the fermionic minus-sign-problem). The sign-problem-free property is a consequence of a fundamental theorem regarding the invariance of the Hamiltonian with respect to an antiunitary symmetry \cite{Wu2005}. For the current model, the Hamiltonian $H(\tau)$ given in Eq. \eqref{FHam} is invariant under the symmetry described by the antiunitary operator $\mathcal{O}=\gamma_{\tsp 1}\mathcal{C}$ defined in terms of the $4\times 4$ Dirac matrix $\gamma_{\tsp 1}$ and the complex conjugation operator $\mathcal{C}$ in the flavor + spin basis $\{\ket{\alpha,s}\}$, such that $\mathcal{O}H(\tau)\mathcal{O}^{-1}=H(\tau)$.

\section{Methodology}\label{sec:III}

\begin{figure}[t]
    \centering
    \begin{tikzpicture}
      \draw (0,0) node[anchor=north, inner sep=0]{
      \includegraphics[width=0.97\columnwidth]{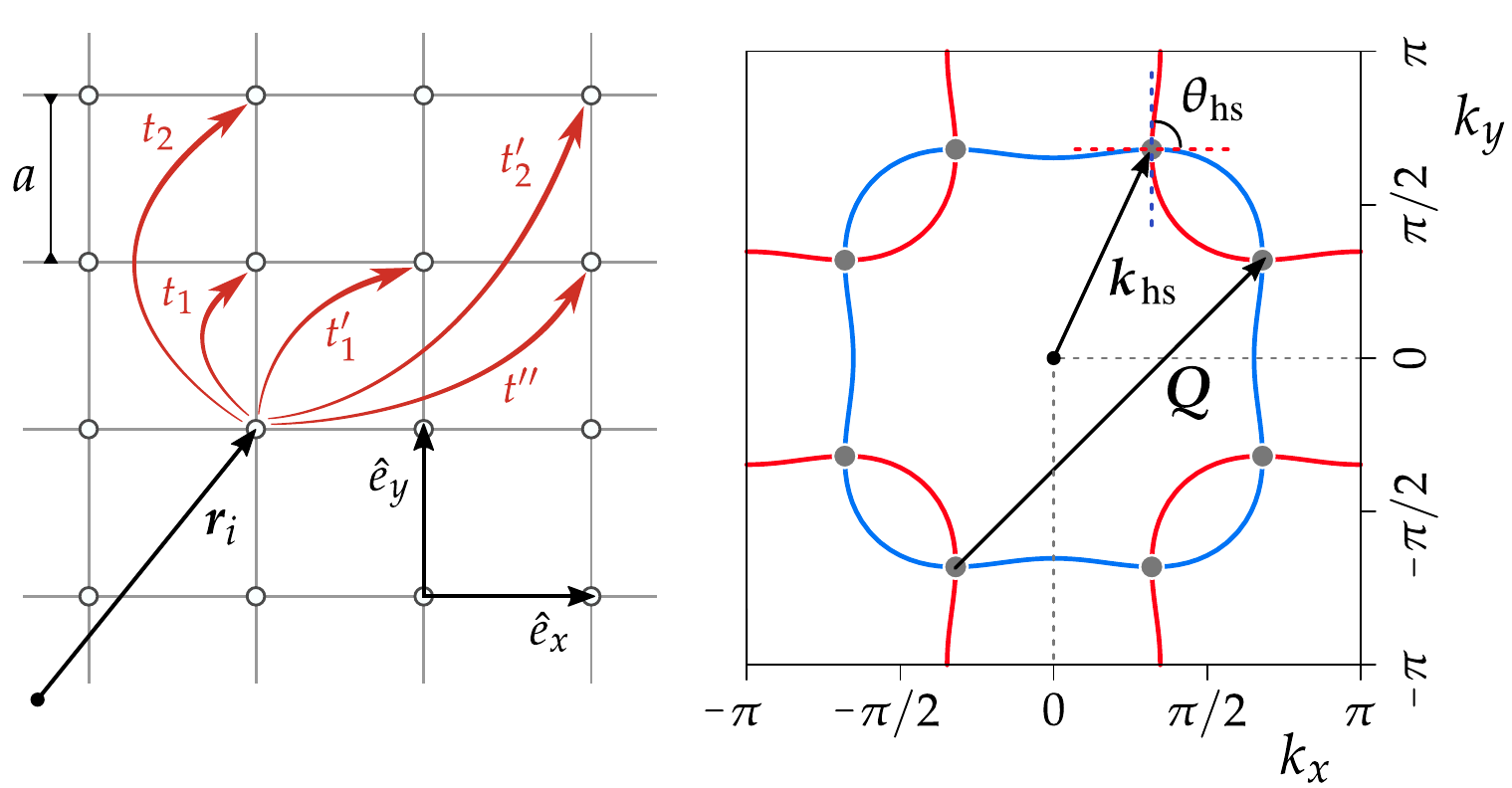}};      
      \draw (-2.35,0.0) node {(a)};
      \draw (+0.40,0.0) node {(b)};
    \end{tikzpicture}
    \caption{(a) The allowed hopping processes considered in the present two-band model. (b) The bare Fermi surfaces in the first Brillouin zone. The components of any wavevector $\bsm{k}=k_{x}\,\hat{e}_{x}+k_{y}\,\hat{e}_{y}$ are measured in units of the inverse lattice constant, the four arcs at the edges correspond to the FS given by $\varepsilon(\bsm{k})=\mu_{\tsp 0}\tsp$, and the closed curve corresponds to the FS given by $\tilde{\varepsilon}(\bsm{k})=\mu_{\tsp 0}\tsp$. The eight large dots at the intersections points mark the hot-spots, where the fermions scatter via the SDW fluctuation coupling field. The red and blue dashed lines are perpendicular to the tangent lines associated with each curve at the highlighted hot-spot momentum $\bsm{k}_{\tsp\text{hs}}\tsp$, which intersect by forming an angle $\theta_{\tsp\text{hs}}\approx \pi/2$ in the figure.}\vspace{-2mm}
    \label{Fig1}
\end{figure}

The determinantal QMC method is a non-perturbative approach that essentially maps the two-dimensional quantum model defined in Eq. \eqref{Lagrangian} onto a (2+1)-dimensional classical model, with the size in imaginary-time direction equal to $\beta$, where the functional integral over the bosonic field is estimated via a Monte-Carlo approach \cite{Grotendorst2002,Assaad2008} (i.e., integrals of the form $\int D\varphi\,\textrm{\scriptsize (\,\dots)}$ are estimated via some importance sampling technique). Here,  we will consider the case of a system described by two degenerate bands, i.e., the hopping parameters are assumed to be the same for both bands: $t^{\tsp\tmath{(1)}}_{\,\tmath{ij}}=t^{\tsp\tmath{(2)}}_{\,\tmath{ij}}\tsp$. The Fig. \ref{Fig1}(a) shows a sketch of the allowed hopping processes in the lattice system, where \cite{RafaelMPTeixeira2022}: $t_{1}=0.6\tsp$, $t_{1}^{\prime}=-0.2\tsp$, $t_{2}=0.12\tsp$, $t^{\prime\prime}=0.02\tsp$, $t_{2}^{\prime}=-0.04\tsp$. We note that, in the non-interacting scenario ($\lambda=0$), the latter parameters result in an energy band with bandwidth $W\approx 5.1$ and dispersion relation $\varepsilon_{\tmath{\bsm{k}}}^{\tsp\tmath{(\alpha)}}=\varepsilon(\bsm{k})$ which yields a FS that bears some resemblance to the experimental FS obtained from angle-resolved photoemission (ARPES) measurements \citep{PRB2007} in the cuprate superconductors. For the choice of chemical potential $\mu_{\tsp 0}=-0.019225\tsp$, the bare Fermi surfaces given by $\varepsilon(\bsm{k})=\mu_{\tsp 0}$ and $\tilde{\varepsilon}(\bsm{k})=\varepsilon(\bsm{k}+\bsm{Q})=\mu_{\tsp 0}$ (i.e., the previous one shifted by the wavevector $\bsm{Q}\tsp$) are plotted within the first Brillouin zone in Fig. \ref{Fig1}(b).

We will be interested in studying how the properties of the system change when the parameters $r$ and $T$ vary, while $\lambda$, $c$, and $u$ remain fixed: $\lambda=4$, $c=2$, and $u=2$. Since in our convention $\lambda^2$ has the dimensions of energy, we have that $\lambda^2/W\approx 3.1$ (i.e., a strong-coupling regime). The system size $L$ in our numerical simulations will be $L=8,10$ and $12$ (with the total number of sites in the lattice given by $N_{s}=L^{2}\tsp$). In this regard, we point out that simulating for even larger lattices takes significantly more CPU time in modern supercomputers using our present QMC code, since the simulation time for a single Monte Carlo step tends to scale (approximately) with a power-law given by $L^{6}$. Furthermore, our choice for the imaginary-time step $\Delta\tau$ varies according to the inverse temperature value. For $\beta>4\tsp$, we set $\Delta\tau=10^{-1}$, otherwise it is given by $\Delta\tau=\beta/M$ with the number of $\tau\,$-$\,$slices ($M$) always close to $40$ (for more details, see also Ref. \cite{RafaelMPTeixeira2022}).

Our investigation will be focused on the transport properties that can be extracted from the time correlation between the imaginary-time uniform current density operator $\vec{\mathcal{J}}(\tau)=\mathcal{J}_{x}(\tau)\,\hat{e}_{x}+\mathcal{J}_{y}(\tau)\,\hat{e}_{y}\,$ ($\,\mathcal{J}_{x}$ and $\mathcal{J}_{y}$ refer to horizontal and vertical components, respectively), where we will only deal with the horizontal component since the system is $C_{4}$ symmetric. The latter is written as $\mathcal{J}_{x}(\tau)=\sum_{\, n}j_{\tsp x}(\bsm{r}_{n},\tau)/L\tsp$, with the operator $j_{\tsp x}(\bsm{r}_{n},\tau)$ expressed in terms of fermionic operators in the Heisenberg representation $c^{\tmath{\tsp(\dagger)}}_{\tsp \alpha,\tsp n,\tsp s}(\tau)=c^{\tmath{\tsp(\dagger)}}_{\tsp\alpha,\tsp s}(\bsm{r}_{n},\tau)$ as follows:
\begin{equation}
    j_{\tsp x}(\bsm{r}_{n},\tau)=-\sum_{\alpha,\tsp s}t_{\tsp x}\!\left[\text{i}\,
    c_{\tsp\alpha,\tsp s}^{\dagger}(\bsm{r}_{n},\tau)\,
    c_{\tsp\alpha,\tsp s}^{\phdg}(\bsm{r}_{n}+\hat{e}_{x}\tsp,\tau)+
    \text{\small H.c.}\right],
\end{equation}
where $n=1,2,...,N_{s}$, and $t_{\tsp x}=t_{1}$ is the nearest neighbor hopping parameter along the unit vector $\hat{e}_{x}$ or $\hat{e}_{y}$ (see Fig. \ref{Fig1}). The imaginary time-ordered current-current correlation function that we will examine corresponds to the grand-canonical ensemble average $\left\langle\tsp\mathcal{T}\,\mathcal{J}_{x}(\tau)\mathcal{J}_{x}(0)\right\rangle\tsp$, which is explicitly calculated as
\vspace{-2mm}
\begin{equation}
    \widetilde{\Lambda}(\tau)=
    \dfrac{1}{N_{s}}\left\langle\tsp \sum_{n,m}
    j_{\tsp x}(\bsm{r}_{n},\tau)\tsp
    j_{\tsp x}(\bsm{r}_{m},0)\!\!\right\rangle
    \equiv\dfrac{1}{N_{s}}\sum_{n,m}
    \widetilde{\Lambda}_{\tsp nm}(\tau).\label{cc_corr}
\end{equation}
Due to the bosonic character of the correlator $\widetilde{\Lambda}(\tau)$, the function $\widetilde{\Lambda}(\tau^{\sprm}+\beta/2)$ is found to be even in the shifted imaginary-time variable $\tau^{\sprm}\in[-\beta/2\,,\beta/2]$, with $\beta/2$ being the half-period value. Numerically, this variable assumes discrete values, i.e. $\tau^{\sprm}_{m}=m\Delta\tau-\beta/2$ ($m$ integer). Hence, when plotting the estimated values for $\widetilde{\Lambda}(\tau^{\sprm}_{m}+\beta/2)$ in terms of $\tau^{\sprm}_{m}$, we observe that the aforementioned parity property is  satisfied within numerical accuracy.

From the calculations required to obtain $\widetilde{\Lambda}(\tau)\,$, one can compute the superfluid density $\rho_{s}$ that provides information about the superconducting state. To this end, one needs the Fourier transform of the current-current correlation function:
\begin{align}
    & \Lambda(\bsm{k},\omega_{l})=
    \sum_{n,m}\,\int_{0}^{\beta}\!d\tau\;\,
    \widetilde{\Lambda}_{\tsp nm}(\tau)\,\delta_{\tsp m,1}\,
    \mbox{e}^{\tsp\text{i}\left(\omega_{l}\tsp\tau\,-\,
    \bsm{k}\,\cdot\,\bsm{R}_{\tsp nm}\right)},\\
    & \Lambda^{L}\equiv\lim_{k_{x}\,\rightarrow\,0}
    \Lambda(\tsp k_{x}\,,\,0)\;\,,\;
    \Lambda^{T}\equiv\lim_{k_{y}\,\rightarrow\,0}
    \Lambda(0\,,\,k_{y}),
\end{align}
where $\omega_{l}=2\tsp l\pi/\beta$ are the bosonic Matsubara frequencies (for integer index $l\tsp$), $\bsm{R}_{\tsp nm}=\bsm{r}_{n}-\tsp\bsm{r}_{m}$ is the lattice vector connecting two sites $n\,$ and $\,m\tsp$, and the Kronecker delta essentially brings $\bsm{r}_{m}$ to the origin of the coordinate system here defined as $\bsm{r}_{\tsp 1}=(0,0)\tsp$. The zero-frequency correlator above is denoted by $\Lambda(\bsm{k},0)=\Lambda(k_{x}\tsp,k_{y})\tsp$, such that the longitudinal ($\Lambda^{L}$) and transverse ($\Lambda^{T}$) limits yield $\rho_{s}=\left(\Lambda^{L}-\Lambda^{T}\right)/4$ (formally, the superfluid density is only given in the limit of $L\rightarrow\infty\tsp$).

The real-frequency conductivity $\sigma(\omega)$ is related to $\widetilde{\Lambda}\left(\tau\right)$ via the following expression
\begin{equation}
    \widetilde{\Lambda}(\tau)=\dfrac{\rho_{q}}{\pi}\int_{0}^{\infty}d\omega\;
    \dfrac{\omega\cosh\left[\left(\beta/2-\tau\right)\omega\right]}
    {\sinh\left(\beta\omega/2\right)}\;\sigma(\omega),
    \label{CC_sigma2}
\end{equation}
where $\rho_{q}=\hbar/e^{2}$ denotes the quantum of resistance. Here, we will measure the inverse conductivity $\rho(\omega)=1/\sigma(\omega)$ (i.e., the real-frequency resistivity) in units of $\rho_{q}\tsp$. In order to extract $\sigma(\omega)$ by inverting the integral above, one could employ the well-known maximum entropy method \citep{PhyRep1996} for analytically continuation of the QMC data related to the current-current correlation function. However, analytical continuation of numerical data is well-known to introduce uncontrollable errors \footnote{We point out that it may be also possible to generalize the determinantal Quantum Monte Carlo method to employ the Kadanoff-Baym contour in order to avoid any need for analytical continuation of the current-current correlation \cite{Eckstein2014}. However, this approach turns out to be computationally more costly and, for this reason, we leave this analysis for a future study (for other systematic ways to circumvent the analytical continuation problem, see also Refs. \cite{LeBlanc2019,Ferrero2020,Ferrero2021})}. Hence, we will employ a proxy for estimating the direct-current (DC) conductivity $\sigma_{\tsp\textrm{\tiny DC}}=\sigma(\omega=0)=1/\rho_{\tsp\textrm{\tiny DC}\tsp}$ (here, $\rho_{\tsp\textrm{\tiny DC}}$ denotes the DC resistivity). A very simple one can be derived from Eq. \eqref{CC_sigma2}. In order to show that, we firstly write the integral kernel in the latter as a $\tau$- and $\omega$-dependent function: $K(\tau,\omega)=\omega\cosh\left[\left(\beta/2-\tau\right)\omega\right]/\sinh\left(\beta\omega/2\right)\tsp$. Then, for $\tau=\beta/2$ (this is effectively the longest possible imaginary-time), we find $K(\beta/2,\omega)=\omega/\sinh\left(\beta\omega/2\right)\tsp$. This function has a full width at half maximum of approximately $\Omega\approx 8.61/\beta\tsp$. Hence, for low enough temperatures, the range of frequencies $[-\Omega,\Omega]$ can be narrow so that $\sigma(\omega)$ can be approximated by its zero-frequency component $\sigma_{\tsp\textrm{\tiny DC}}$ if its low-frequency character is preserved when $\left|\omega\right|<\Omega\tsp$. Then \citep{PRB1996}
\begin{equation}
    \widetilde{\Lambda}\left(\beta/2\right)\approx\!
    \left[\,\int_{0}^{\infty}\dfrac{d\omega}{\pi}\;K(\beta/2,\omega)\,\right]\!
    \sigma_{\tsp\textrm{\tiny DC}}=\dfrac{\pi}
    {\beta^{\tsp 2}\rho_{\tsp\textrm{\tiny DC}}}\;.
    \label{prx0}
\end{equation}
Since $\widetilde{\Lambda}(\tau)$ satisfies the relation $g(\beta-\tau)=g(\tau)$ (with $g$ denoting a correlator of bosonic character), the ``long-time'' behavior that the current-current correlator develops at times close to the half period value $\beta/2$ can be used to estimate the DC resistivity via the proxy: $\rho^{\,\textrm{pr,1}}_{\tsp\textrm{\tiny DC}}=\pi/[\tsp\beta^{2}\widetilde{\Lambda}\left(\beta/2\right)]\tsp$. In Ref. \citep{PNAS2017}, it was shown for an Ising-nematic quantum critical model with spin-1/2 itinerant electrons that a simple proxy like the latter is not enough to capture the low-frequency character of $\sigma(\omega)$. A more suitable proxy is another one that involves more details on the ``long-time'' behavior of $\widetilde{\Lambda}(\tau)\tsp$, and it can be derived by noticing that the second derivative of the correlator $\widetilde{\Lambda}^{\prime\prime}(\tau)=\partial_{\tau}^{\tsp 2}\tsp\widetilde{\Lambda}(\tau)$ and $\sigma(\omega)$ are connected via an integral relation involving the kernel function $K^{\prime\prime}(\tau,\omega)=\partial_{\tau}^{\tsp 2}\tsp K(\tau,\omega)=\omega^{\tsp 2}\tsp K(\tau,\omega)\tsp$. For $\tau=\beta/2\tsp$, this function peaks at $\omega=6/\beta$ (assuming positive frequencies) and decays exponentially for $\omega>\Omega^{\prime}$ where $\Omega^{\prime}\approx 11.04/\beta$ is the frequency associated with the half maximum. Then, if the range of frequencies $[-\Omega^{\prime},\Omega^{\prime}]$ is narrow enough so that $\sigma(\omega)$ can be approximated by $\sigma_{\tsp\textrm{\tiny DC}}$ just like before, one finds:
\begin{equation}
    \widetilde{\Lambda}^{\prime\prime}(\beta/2)
    \approx\!\left[\,\int_{0}^{\infty}\dfrac{d\omega}{\pi}\;
    K^{\prime\prime}(\beta/2,\omega)\,\right]\!
    \sigma_{\tsp\textrm{\tiny DC}}=\dfrac{2\pi^{\tsp 3}}{
    \beta^{\tsp 4}\tsp\rho{\tsp\textrm{\tiny DC}}}.
    \label{prx1}
\end{equation}
Now, from Eq. \eqref{prx0}, we have $[\tsp\pi\tsp\widetilde{\Lambda}(\beta/2)]^{n}\approx\pi^{\tsp 2n}/(\beta^{\tsp 2n}\rho_{\tsp\textrm{\tiny DC}}^{\tsp n})$ with $n>1$ being an integer number. Then, by combining the latter expression with the relation between $\widetilde{\Lambda}^{\prime\prime}(\beta/2)$ and $\rho{\tsp\textrm{\tiny DC}}$ given by Eq. \eqref{prx1}, we can write that
\vspace{-1mm}
\begin{equation}
    \dfrac{\pi\tsp\widetilde{\Lambda}^{\prime\prime}(\beta/2)}{
    2[\tsp\pi\tsp\widetilde{\Lambda}(\beta/2)]^{n}}\approx
    \rho_{\tsp\textrm{\tiny DC}}^{\tsp n-1}\!
    \left(\dfrac{\beta}{\pi}\right)^{\!2(n-2)}\;.
\end{equation}
Thus, for $n=2\tsp$, a proxy for the DC resistivity involving both $\widetilde{\Lambda}(\beta/2)$ and $\widetilde{\Lambda}^{\prime\prime}(\beta/2)$ is given (in units of $\rho_{q}$) by
\vspace{-1mm}
\begin{equation}
    \rho^{\,\textrm{pr\tsp,\tsp 2}}_{\tsp\textrm{\tiny DC}}=
    \dfrac{1}{2\pi}\,\dfrac{\widetilde{\Lambda}^{\prime\prime}(\beta/2)}
    {[\widetilde{\Lambda}(\beta/2)]^{2}},
    \label{prx2}
\end{equation}
whereas, for $n=3\tsp$, one can define another proxy given by
\vspace{-1mm}
\begin{equation}
    \rho^{\,\textrm{pr\tsp,\tsp 3}}_{\tsp\textrm{\tiny DC}}=
    \dfrac{1}{\beta}\sqrt{
    \dfrac{\widetilde{\Lambda}^{\prime\prime}(\beta/2)}{
    2[\tsp\widetilde{\Lambda}(\beta/2)]^{3}}}\;.
    \label{prx3}
\end{equation}
Henceforth, we will focus only on the proxy given by $\rho^{\,\textrm{pr\tsp,\tsp 2}}_{\tsp\textrm{\tiny DC}}$, since we verified that the proxies of Eqs. \eqref{prx2} and \eqref{prx3} yield qualitatively similar results for the resistivity in the present model. For this reason, we will refer to the proxy of Eq. \eqref{prx2} as simply $\rho^{\,\textrm{\tiny proxy}}_{\tsp\textrm{\tiny DC}}$. This latter proxy was also shown to yield excellent results when compared to the analytically continued QMC data for the DC resistivity, e.g., of the 2D Hubbard model \citep{Sci2019}.

In our investigation of transport properties of the model, we use the fact that $\widetilde{\Lambda}(\tau)$ has bosonic character so the polynomial function $F(\tau)=\sum_{\,n\tsp=\tsp 1}^{\,2}b_{\tsp 2n}(\tau-\beta/2)^{2n}$ can be fitted to the QMC data for the shifted current-current correlator 
$\widehat{\Lambda}(\tau_{m})=\widetilde{\Lambda}(\tau_{m})-\widetilde{\Lambda}(\beta/2)$. This fitting procedure captures the ``long-time'' behavior of the latter while also filtering out the fluctuations in the estimated values when $m\simeq M/2\tsp$. In our implementation, we perform successive fits of the function $F(\tau)$ using data sets with increasing length $2p+1$ containing the estimated values for $\widehat{\Lambda}(\tau_{m})$ such that $m=M_{\tsp h}-p,M_{\tsp h}-p+1,\,\dots\,,M_{\tsp h}+p$, where $M_{\tsp h}=M/2$ is the central-point index ($\tau_{\tmath{M/2}}=\beta/2$) and $1<p<M_{h}\tsp$. We choose the data set for which the fitting function $F(\tau)$ better describes the QMC data near the central-point while also giving a reasonable fit of the data for shorter-times (i.e., far from the central-point). Then, our QMC data for the correlator at imaginary-times $\tau\simeq\beta/2$ is replaced by this function: $\widehat{\Lambda}(\tau_{m})\rightarrow F(\tau_{m})=\widehat{\Lambda}_{\,\text{fit}}(\tau_{m})\tsp$. Thus, when estimating the proxy for the DC resistivity given in Eq. \eqref{prx2}, the numerator will be replaced by the coefficient $b_{2}\tsp$, i.e., $\widetilde{\Lambda}^{\prime\prime}(\beta/2)\rightarrow\widehat{\Lambda}_{\,\text{fit}}^{\prime\prime}(\beta/2)=b_{2}$.

\begin{figure*}[!t]
    \centering
    \includegraphics[width=2.05\columnwidth]{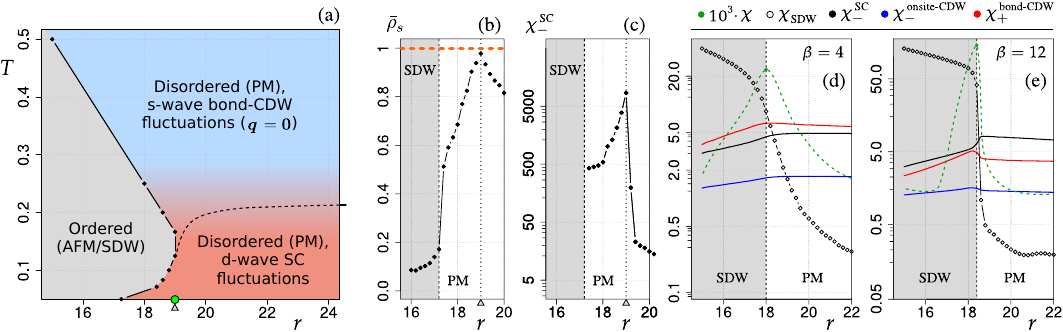}
    \caption{(a) Phase diagram of temperature $T$ versus tuning parameter $r$ for the two-dimensional $Z_2$ spin-fermion model obtained in this work. (b) Plot of the rescaled superfluid density $\bar{\rho}_{s}$. (c) Log-linear plot of the $d$-wave SC susceptibility $\chi_{\tsp -}^{\text{\smaller\,SC}}$ as a function of $r$ for $\beta=18$ in a narrow range where both quantities peak (maximum value indicated by the triangular mark in the $r$-axis where $r_{\text{\smaller SC}}=19\tsp$). In (d) and (e), log-linear plots of the most relevant susceptibilities calculated for the present model, namely: finite-lattice susceptibility $\chi$ (fitted values given by the rescaled dashed curve), SDW susceptibility $\chi^{\vspace{1mm}}_{\tsp\text{\smaller SDW}}\tsp$, $d$-wave SC susceptibility $\chi_{\tsp -}^{\text{\smaller\,SC}}\tsp$, $d$-wave onsite-CDW susceptibility $\chi^{\text{\smaller\,onsite-CDW}}_{\tsp -}$ for $\bsm{k}=\bsm{Q}\tsp$, $s$-wave bond-CDW susceptibility $\chi^{\text{\smaller\,bond-CDW}}_{\tsp +}(\bsm{k})$ with $\bsm{k}\approx\bsm{0}$ in (d) and $|\bsm{k}|>0$ in (e). The green point in (a) corresponds to the peak of the $d$-wave superconducting dome as signaled by the quantities in the plots (b) and (c). The competition between SC and CO fluctuations is shown schematically by the color gradient in the PM region of the diagram in (a). In the same plot, these two competing fluctuations (the most important ones in the PM phase) are almost equal in magnitude (absolute value of the susceptibility) in the close vicinity of the dashed curved line, which was estimated via linear interpolation of the QMC data in the $T$ domain. In (c), we only showed the data for the PM phase where the statistical errors for $\chi_{\tsp -}^{\,\text{\smaller SC}}$ were reasonably small, and in (b) we found that $\bar{\rho}_{s}-1$ at $r_{\text{\smaller SC}}=19$ (see the triangular mark and the orange horizontal dashed line) is smaller than the associated error. The diagram was estimated via the analysis of the QMC data from simulations of an $8\times 8$ system. SDW, SC, CDW, and PM refer to spin density-wave, superconductivity, charge density-wave, and paramagnetic, respectively.}
    \label{L8_diagram}
\end{figure*}

\section{QMC Results}\label{sec:IV}

\subsection{Phase diagram}

In order to estimate the magnetic phase diagram for the present model, we examine the momentum resolved bosonic spin-density-wave (SDW) susceptibility for a commensurate SDW order (i.e., at the wavevector $\bsm{Q}\tsp$), which is calculated in terms of the grand-canonical ensemble average for the finite-system magnetization \cite{Cuccoli1995} $\mathcal{M}(\varphi)=1/(M N_{s})\sum_{\, i,m}\varphi_{i,m}$ as $\chi^{\vspace{1mm}}_{\tsp\text{\smaller SDW}}=\beta N_{s}\langle\mathcal{M}^{\tsp 2}(\varphi)\rangle\tsp$ (notice that $\mathcal{M}(\varphi)$ is simply the average over all sites and imaginary-time slices for a sampled configuration $\{\varphi_{i,m}\}\tsp$). For a fixed inverse temperature $\beta\tsp$, $\chi^{\vspace{1mm}}_{\tsp\text{\smaller SDW}}$ is strongly enhanced as the tuning parameter $r$ is varied within a certain range of values $[15,r_{c}(\beta)]$ with $r_{c}(\beta)$ being a temperature dependent critical value (see Fig. \ref{L8_diagram}).

Through the analysis of quantities such as the local moment \cite{Santos2003}, the average double- and single-site occupancy, and also the averages of both fermionic and bosonic energies (i.e., the grand-canonical ensemble averages of the Hamiltonian in Eq. \eqref{FHam} and of the Lagrangian terms in Eq. \eqref{BLag}), one infers that the system is in the AFM/SDW phase for $r<r_{c}(\beta)\tsp$, and that the paramagnetic (PM) phase, established when $r>r_{c}(\beta)$, is characterized by an increase in the degree of itinerancy of the fermions and disordered sampled configurations of the bosonic field. For instance, the average site occupancy is calculated as $\langle n\rangle=1/(2N_{s})\sum_{\,i\,}\langle n_{\tsp i}\rangle$ with $n_{\tsp i}=\sum_{\,\alpha,\tsp s}\resizebox{!}{8.2pt}{$c_{\alpha,\tsp i,\tsp s}^{\dagger}\, c_{\alpha,\tsp i,\tsp s}^{\phdg}$}$ (total occupation operator for the $i$-th site) which can be related to the doping parameter $\delta=\langle n \rangle-1\tsp$, such that $\langle n \rangle=1$ (half-filling) implies $\delta=0$. In the present model, $\delta$ is very small when $r\ll r_{c}(\beta)$ and it increases with $r$ as the latter is tuned across the ``critical region'' $r\sim r_{c}$ (more on that in the next section). For $r<r_{c}(\beta)\tsp$, we find that $\chi^{\vspace{1mm}}_{\tsp\text{\smaller SDW}}(r,\beta)\simeq\xi(\beta)\,\text{e}^{-b\tsp r}\tsp$, where $\xi(\beta)\simeq a_{\smath{\tsp 0}}+a\,\text{log}(1/T)$ with $a_{\smath{0}}\tsp$, $a$ and $b$ being positive real numbers. As $r$ is increased beyond $r_{c}(\beta)\tsp$, we find that $\chi^{\vspace{1mm}}_{\tsp\text{\smaller SDW}}$ is strongly suppressed as it tends to decrease following closely a $1/r$ power law (for fixed $\beta\tsp$) in the PM phase. The value of $r_{c}(\beta)$ can be determined from the QMC data for a fixed temperature by examining the behavior of many order-parameter susceptibilities, since a noticeable increase or decrease (mainly at low temperatures) in the numerical values is found when $r$ is tuned across the critical region in between the ordered and disordered phases. Particularly, {the finite-lattice susceptibility} \cite{Landau2014} (which is proportional to the variance corresponding to the measurements of $\mathcal{M}(\varphi)$) $\chi=\chi^{\vspace{1mm}}_{\tsp\text{\smaller SDW}}\!-\!\beta N_{s}\langle|\mathcal{M}(\varphi)|\rangle^{2}$ is very useful in this regard as it shows a prominent peak at $r_{c}(\beta)\tsp$, such that the latter can be estimated (for a certain system size $L\tsp$) by looking at where $\chi$ is maximum (an example of this is shown in the Appendix). In this way, we obtained the AFM/SDW phase boundary displayed in Fig. \ref{L8_diagram}(a).

In order to extract the information about the SC state, we followed the approach of Ref. \cite{Paiva2004} for the determination of the SC critical temperature $T_{c}$ of a state of Berezinskii-Kosterlitz-Thouless (BKT) character, which is associated with a superfluid density (for details about the estimation of such quantity, see the Refs. \cite{Santos2003,Paiva2004,Foley2019}) that exceeds the universal BKT value $C_{\tsp\text{\smaller BKT}}(T)=2T/\pi\tsp$. 
Hence, for each fixed temperature value $T\tsp$, we mapped out a range of the tuning parameter where the quantity $\bar{\rho}_{s}(T)=\rho_{s}(T)/C_{\tsp\text{\smaller BKT}}(T)$ is greater than unity. In Fig. \ref{L8_diagram}(b), the QMC results obtained from the simulations at $\beta=18$ are shown for $\bar{\rho}_{s}\tsp$. For the present model, we find that $\bar{\rho}_{s}$ is very close to unity (but still slightly below this value) for a single point in the diagram: $\beta=18$ (or $T\approx 0.057$) and $r=r_{\text{\smaller SC}}\tsp$. This peak at $r=r_{\text{\smaller SC}}$ is expected to increase for $\beta>18\tsp$, such that the shape of a SC dome might be revealed at lower temperatures. By fixing $r$ at such a peak value and performing a linear fit of the quantity $\ln\bar{\rho}_{s}(T)$ for $\beta\in[6,18]$, we find that $\ln\bar{\rho}_{s}(T)=a+b\tsp T$, where the coefficients are given by $a=1.43\pm 0.06$ and $b=-27.5\pm 0.6\tsp$. Thus, if we extrapolate this fit of the QMC data to $\beta>18\tsp$, the SC transition will likely be found at the inverse temperature $\beta_{c}=19.2\pm 0.9$ (i.e., $T_{c}\approx 0.052\tsp$). We point out that this critical temperature agrees within numerical accuracy with the general theoretical formula derived from Eliashberg theory for the $O(N)$ spin-fermion model \cite{Berg_cond-mat,Wang:2016tr}, which predicts that $\beta_{c}\approx 19.7$ for the present model. As a result, although the Eliashberg formula in principle assumes a weak spin-fermion coupling, our results reaffirm that it holds even in a stronger coupling regime in qualitative agreement with the conclusions of Ref. \cite{Wang:2016tr}.

The nature of the SC state can be extracted from the quantity $\Delta\chi^{\,\text{\smaller SC}}=\chi_{\tsp +}^{\,\text{\smaller SC}}-\chi_{\tsp -}^{\,\text{\smaller SC}}$ defined as the difference between the $s$-wave ($\eta=+1$) form and the $d$-wave ($\eta=-1$) form of the uniform ($\tsp\bsm{k}=\bsm{0}\tsp$) superconducting susceptibility\cite{Schattner:2016dw,Lee2017}
\begin{equation}
    \chi_{\tsp\eta}^{\,\text{\smaller SC}}(\bsm{k})=
    \int_{0}^{\beta}\dfrac{d\tau}{N_{s}}\,\sum_{i,j}^{
    \text{\resizebox{!}{4.4pt}{$N_{s\vphantom{x_{x}}}$}}}
    \langle P_{\eta}^{\tsp\dagger}(\bsm{r}_{i},\tau)
    P_{\eta}(\bsm{r}_{j}\tsp,0)\rangle\,e^{\,\text{i}\hspace{0.5pt}\bsm{k}\,
    \cdot\,(\bsm{r}_{i}-\bsm{r}_{j})},\label{chi_PDW}
\end{equation}
where $P_{\eta}(\bsm{r}_{i},\tau)$ is an auxiliary operator associated with the SC order (in the $\bsm{r}$-basis representation) defined as follows:
\begin{align}    
    & \!\!P_{\eta}(\bsm{r}_{i},\tau)=
    \sum_{\alpha \,=\, 1^{\vphantom{X}}}^{2_{\vphantom{X}}}
    \!\eta^{\tsp\alpha-1}\!\left[
    c_{\tsp\alpha,\tsp i,\tsp\uparrow  }^{\dagger}(\tau)\tsp
    c_{\tsp\alpha,\tsp i,\tsp\downarrow}^{\dagger}(\tau)-
    c_{\tsp\alpha,\tsp i,\tsp\downarrow}^{\dagger}(\tau)\tsp
    c_{\tsp\alpha,\tsp i,\tsp\uparrow  }^{\dagger}(\tau)\right]
    \nonumber\\[1mm]
    & \!\!\hphantom{P_{\eta}(\bsm{r}_{i},\tau)}=\tsp
    2\left[c_{\tsp 1,\tsp i,\tsp\uparrow  }^{\dagger}(\tau)\tsp
           c_{\tsp 1,\tsp i,\tsp\downarrow}^{\dagger}(\tau)+
    \eta\, c_{\tsp 2,\tsp i,\tsp\uparrow  }^{\dagger}(\tau)\tsp
           c_{\tsp 2,\tsp i,\tsp\downarrow}^{\dagger}(\tau)\right].
    \label{PDW_operator}
\end{align}
As for an operator associated with charge order (CO), we obtain the susceptibility $\chi_{\tsp\eta}^{\,\text{\smaller CO}}(\bsm{k})$ by means of an imaginary-time integral analogous to the previous one. We examine CO of two types: onsite-CDW order ($\tsp\chi^{\,\text{\smaller onsite-CDW}}_{\tsp\eta}\tsp$) and bond-CDW order ($\tsp\chi^{\,\text{\smaller bond-CDW}}_{\tsp\eta}\tsp$). {They are given by the following expressions:} 
\begin{equation}
    \begin{aligned}
        & \!\!\!\chi^{\,\text{\smaller onsite-CDW}}_{\tsp\eta}(\bsm{k})=
        \!\int_{0}^{\beta}\dfrac{d\tau}{N_{s}}\,\sum_{i,j}^{
        \text{\resizebox{!}{4.4pt}{$N_{s\vphantom{x_{x}}}$}}}
        \langle C_{\eta}^{\tsp\dagger}(\bsm{r}_{i}\tsp,\tau)
        C_{\eta}(\bsm{r}_{j}\tsp,0)\rangle\,e^{\,\text{i}\hspace{0.5pt}\bsm{k}\,
        \cdot\,(\bsm{r}_{i}-\bsm{r}_{j})},\\[-0.4mm]
        & \!\!\!\chi^{\,\text{\smaller bond-CDW}}_{\tsp\eta}(\bsm{k})=
        \!\int_{0}^{\beta}\dfrac{d\tau}{N_{s}}\,\sum_{i,j}^{
        \text{\resizebox{!}{4.4pt}{$N_{s\vphantom{x_{x}}}$}}}
        \langle B_{\eta}^{\tsp\dagger}(\bsm{r}_{i}\tsp,\tau)
        B_{\eta}(\bsm{r}_{j}\tsp,0)\rangle\,e^{\,\text{i}\hspace{0.5pt}\bsm{k}\,
        \cdot\,(\bsm{r}_{i}-\bsm{r}_{j})},
    \end{aligned}
    \label{chi_CDW}
\end{equation}
where $\eta=\pm1$ are again associated with the $s$ and $d$-wave forms. In the calculation of $\chi^{\,\text{\smaller bond-CDW}}_{\tsp\eta}\tsp$, we consider only nearest-neighbor bonds \cite{Lee2017}. The auxiliary operators for these two CDW orders can be expressed as $C_{\eta}(\bsm{r}_{i})=\sum_{\,s}C_{\eta}^{\,s}(\bsm{r}_{i})$ and $B_{\eta}(\bsm{r}_{i})=\sum_{\,s}B_{\eta}^{\,s}(\bsm{r}_{i})$, where, for onsite- and bond-CDW orders respectively, the sum over the spin index $s=\uparrow,\downarrow$ involves the following imaginary-time dependent operators (we omit $\tau$ for compactness):
\begin{align}
    C_{\eta}^{\,s}(\bsm{r}_{i}) & \!=\!\left[
    c_{\tsp 1,\tsp s}^{\dagger}(\bsm{r}_{i})\tsp 
    c_{\tsp 1,\tsp s}^{\phdg  }(\bsm{r}_{i})+\eta\,
    c_{\tsp 2,\tsp s}^{\dagger}(\bsm{r}_{i})\tsp 
    c_{\tsp 2,\tsp s}^{\phdg  }(\bsm{r}_{i})\right],\nonumber\\
    \label{CDW_operator}\\[-4.4mm] 
    B_{\eta}^{\,s}(\bsm{r}_{i}) & \!=\!\left[
    c_{\tsp 1,\tsp s}^{\dagger}(\bsm{r}_{i})\tsp
    c_{\tsp 1,\tsp s}^{\phdg  }(\bsm{r}_{i}+\hat{e}_{x})+
    c_{\tsp 1,\tsp s}^{\dagger}(\bsm{r}_{i})\tsp
    c_{\tsp 1,\tsp s}^{\phdg  }(\bsm{r}_{i}-\hat{e}_{x})\tsp\right]\!+
    \nonumber\\
    & \hspace{1.6mm}\eta\!\left[
    c_{\tsp 2,\tsp s}^{\dagger}(\bsm{r}_{i})\tsp
    c_{\tsp 2,\tsp s}^{\phdg  }(\bsm{r}_{i}+\hat{e}_{y})+
    c_{\tsp 2,\tsp s}^{\dagger}(\bsm{r}_{i})\tsp
    c_{\tsp 2,\tsp s}^{\phdg  }(\bsm{r}_{i}-\hat{e}_{y})\right]\!+
    \textrm{H.c..}\nonumber
\end{align}
We note that the $d$-wave symmetry of the set of operators $\{P_{-1}^{\vphantom{\,s}},C_{-1}^{\,s},B_{-1}^{\,s}\}$ can be understood by considering a system composed of non-degenerate bands defined in such a way that $\pi/2$ rotations in momentum space transform one band into the other. This can be achieved by slightly deforming the initially degenerate bands by assuming horizontal ($x$) and vertical ($y$) nearest neighbors hopping parameters given by: $t_{1,x}^{\tsp\tmath{(\alpha)}}=t_{1}+(-1)^{\tsp\alpha-1}\Delta t$ and $t_{1,y}^{\tsp\tmath{(\alpha)}}=t_{1}+(-1)^{\tsp\alpha}\Delta t\tsp$ (where $\Delta t>0$ measures the magnitude of the deformation). In this scenario, the $\pi/2$ rotations are equivalent to exchanging the band indices. Applying this transformation to the auxiliary operators from Eqs. \eqref{PDW_operator} and \eqref{CDW_operator} changes their signs, as expected. We argue that these operators remain the same in the limit of degenerate bands (i.e., $\Delta t\rightarrow 0$), since the properties of the model should not be sensitive to small modifications of the hopping parameters.

\begin{figure}[!t]
    \centering
    \includegraphics[width=1.0\columnwidth]{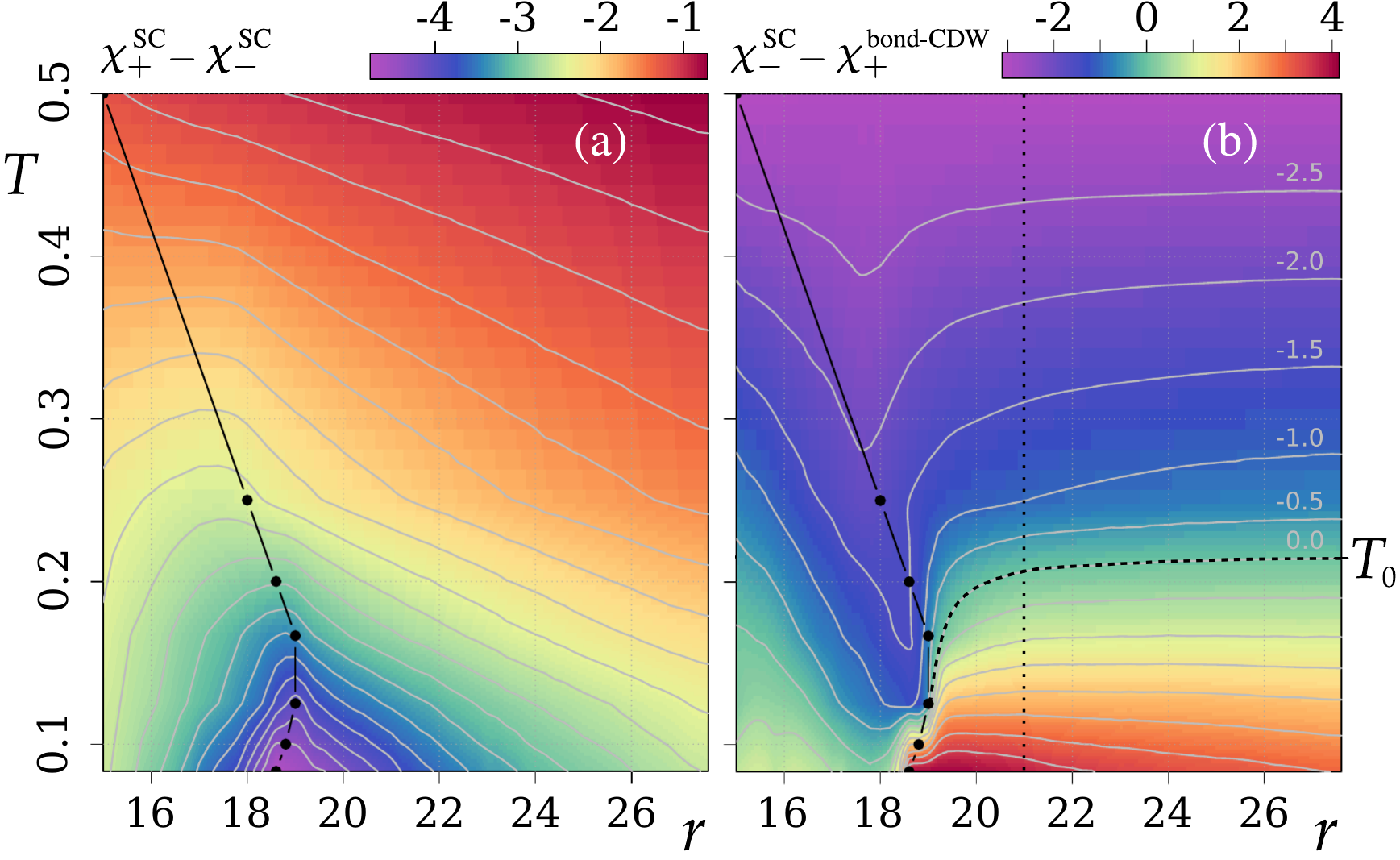}
    \caption{Diagram view ($T$-$\tsp r$ plot) showing the temperature dependence of the susceptibilities: (a) $\Delta\chi^{\,\text{\smaller SC}}$ and (b) $\Delta\chi^{\,\text{\smaller SC\tsp/\tsp CO}}$ for different fixed $r$ parameter values. The curved dashed line in (b) follows the contour corresponding to $\Delta\chi^{\,\text{\smaller SC\tsp/\tsp CO}}=0\tsp$, and the vertical dotted line at $r=21$ roughly divides the diagram into two parts: AFM region and PM region. In the latter, $\Delta\chi^{\,\text{\smaller SC\tsp/\tsp CO}}$ is weakly dependent on $r\tsp$, so that there is an approximate temperature value $T_{\tmath{0}}\approx 0.2125$ (estimated from the linear interpolated data points and it is indicated in the right-edge of the plot (b)) for which CO fluctuations of the type $s$-wave bond-CDW dominate in the PM phase if $T>T_{\tmath{0}}\tsp$, while $d$-wave SC fluctuations dominate at low temperatures $T<T_{\tmath{0}}\tsp$.}
    \label{Delta_chi}
\end{figure}

In Fig. \ref{Delta_chi}, the dependence with $r$ and $T$ of the difference of susceptibilities $\Delta\chi^{\,\text{\smaller SC}}$ and $\Delta\chi^{\,\text{\smaller SC\tsp/\tsp CO}}=\chi_{\tsp -}^{\,\text{\smaller SC}}-\chi^{\,\text{\smaller bond-CDW}}_{\tsp +}$ are shown for $\beta\leq 12\tsp$. The color-coded plots in the figure represent the results of linear interpolations of the QMC data in the $T$-domain (the interpolated data provide a much better visualization of the behavior of the plotted quantity along the whole diagram region since the resolution in the both directions are similar). The plot in Fig. \ref{Delta_chi}(a) reveals that SC fluctuations of $d$-wave character are always stronger than those of $s$-wave character (i.e., $\tsp\Delta\chi^{\,\text{\smaller SC}}<0\tsp$), with $\chi^{\text{\tiny\,SC}}_{-}$ being more strongly enhanced relatively to $\chi^{\text{\tiny\,SC}}_{+}$ at low temperatures and in the vicinity of the magnetic transition $r\sim r_{c}(\beta)\tsp$, which is an expected behavior since a model with a bosonic order parameter of higher dimensionality \cite{Sci2012,Schattner:2016dw,Berg_2019,Berg_cond-mat} yields similar results. In the same plot, the ratio $\chi_{\tsp -}^{\,\text{\smaller SC}}/\chi_{\tsp +}^{\,\text{\smaller SC}}$ is slightly larger than $2.4$ at $r_{\text{\smaller SC}}=19$ and $\beta=12\tsp$, which is the $r$ value where the quantity $\left|\tsp\Delta\chi^{\,\text{\smaller SC}}\right|$ is maximum and also the one associated with the peak of the $d$-wave SC dome indicated previously in Figs. \ref{L8_diagram}(a)-(c). When analyzing $\chi_{\tsp -}^{\,\text{\smaller SC}}$ in the PM phase, we noticed that it increases monotonically with $1/T$ at a rate that is gradually weakened as the positive difference $\Delta r_{c}=\tsp r-r_{c}(\beta)>0$ increases and, for $r<19\tsp$, these SC fluctuations become saturated when $T$ decreases below the magnetic transition threshold.

Moreover, in the disordered phase, we found that uniform ($\tsp\bsm{k}=\bsm{0}\tsp$) CDW fluctuations of bond-type with $s$-wave character compete with the $d$-wave SC fluctuations (this is why we examined the susceptibility given by $\Delta\chi^{\,\text{\smaller SC\tsp/\tsp CO}}$ which is plotted in Fig. \ref{Delta_chi}), i.e., for $r>r_{c}(\beta)\tsp$, there are two main regions in the diagram: $\Delta\chi^{\,\text{\smaller SC\tsp/\tsp CO}}<0$ (CO fluctuations dominate), $\Delta\chi^{\,\text{\smaller SC\tsp/\tsp CO}}>0\tsp$ (SC fluctuations dominate). For $\beta=4\tsp$, $\chi^{\,\text{\smaller bond-CDW}}_{\tsp +}>\chi_{\tsp -}^{\,\text{\smaller SC}}$ in the whole $r$ parameter range as shown in the plot of Fig. \ref{L8_diagram}(d), while in Fig. \ref{L8_diagram}(e) we see that for $\beta=12$ the SC fluctuations become dominant: $\chi_{\tsp -}^{\,\text{\smaller SC}}>\chi^{\,\text{\smaller bond-CDW}}_{\tsp +}\tsp$. In Fig. \ref{Delta_chi}(b), the regions where CO or SC fluctuations dominate emerge clearly in the diagram. As explained in the caption of the same figure, for $r\gtrsim 21$ (see the vertical dotted line), we can find an approximate temperature value $T_{\tmath{0}}$ so a dashed line given by $T=T_{\tmath{0}}$ separates these two regions in the PM phase. This is consistent with the result shown in Fig. \ref{L8_diagram}(a). For $r\lesssim 21\tsp$, we see that the region where the $\Delta\chi^{\,\text{\smaller SC\tsp/\tsp CO}}<0$ extends along the AFM/SDW phase boundary down to temperatures $T\simeq 0.125$ below $T_{\tmath{0}}\tsp$. Thus, at low temperatures $T\lesssim 0.1$, SDW order competes mainly with the increasing $d$-wave SC fluctuations in the close vicinity of the magnetic ordered phase region (i.e., near the AFM-PM transition), while at temperatures $T\gtrsim 0.125\tsp$, the main type of fluctuations that competes with the SDW order are of $s$-wave bond-CDW-type. {Interestingly, this type of CO fluctuation can be associated with finite ordering wavevectors when $\beta\gtrsim 6\tsp$. In the case of onsite-CDW order, the results indicate that the ordering wavevector coincides with $\bsm{Q}$ for system sizes up to $L=12\tsp$, since the susceptibility $\chi^{\text{\smaller\,onsite-CDW}}_{\tsp\eta}$ was always found to peak at $\bsm{k}=\bsm{Q}\tsp$.}

\subsection{Resistivity of the normal phase}

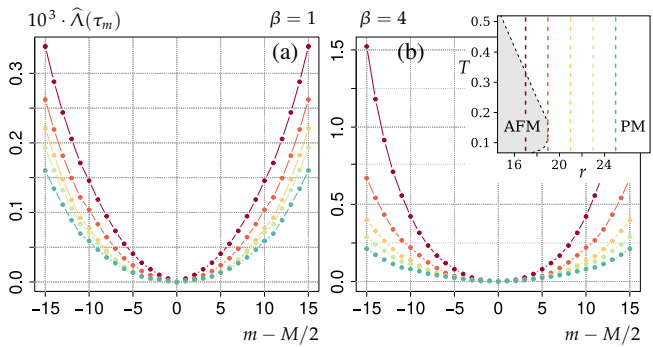
\begin{figure}[!t]
    \centering
    \begin{tikzpicture}
        \draw (0,0) node[anchor=north, inner sep=0]{
        \includegraphics[width=1.0\columnwidth]{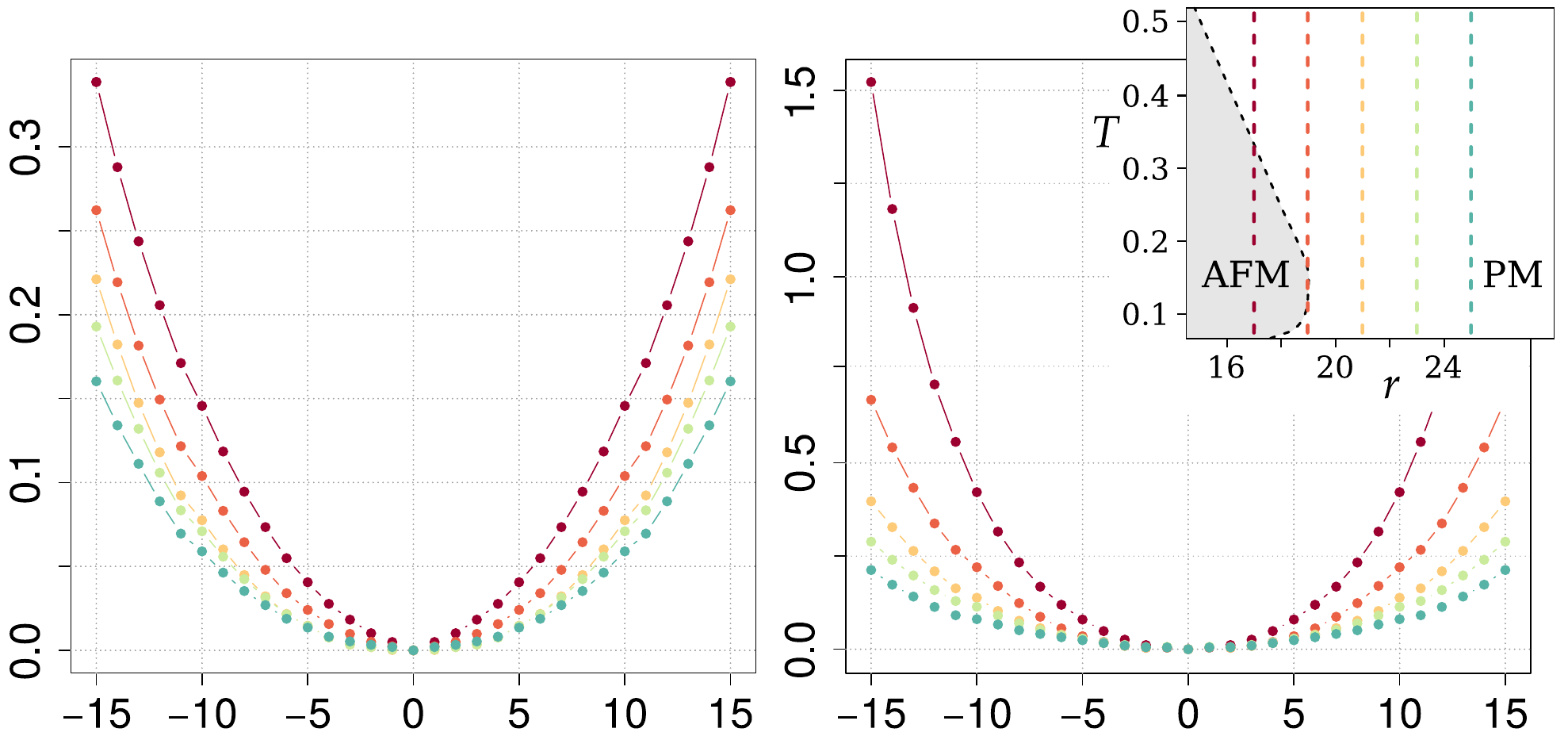}};
        \draw (-0.60,-0.55) node {(a)};
        \draw (+1.07,-0.55) node {(b)};
        \draw (-0.66,-4.3) node {$\smath{m-M/2}$};
        \draw (+3.62,-4.3) node {$\smath{m-M/2}$};
        \draw (-3.4,-0.1) node {
        $\smath{10^{\tsp 3}\cdot\widehat{\Lambda}(\tau_{m})}$};
        \draw (-0.5,-0.1) node {$\smath{\beta=1}$};
        \draw (+0.7,-0.1) node {$\smath{\beta=4}$};
    \end{tikzpicture}
    \caption{Imaginary time behavior of the current-current correlator $\widetilde{\Lambda}(\tau_{m})$ for inverse temperatures (a) $\beta=1$ and (b) $\beta=4$ (results from QMC simulations of a model with lattice size $L=12\tsp$). The estimated valued for $\widehat{\Lambda}(\tau_{m})=\widetilde{\Lambda}(\tau_{m})-\widetilde{\Lambda}(\beta/2)$ are shown for increasing $r$ values (see inset plot) and $m=M/2-p,\dots,M/2+p$ with $p=15\tsp$. The $r$ values associated with each data set is color-coded according to the inset plot in the top-right corner, where the vertical dashed lines give a reference for the proximity of $r$ to the AFM/SDW phase boundary. Here, we assumed that the function $f(\tau_{m}-\beta/2)=\widehat{\Lambda}(\tau_{m})$ is an even function of the discretized imaginary time $\tau_{m}=m\tsp\Delta\tau\tsp$, with $m=1,2,\dots,M$ and $M=40\tsp$.}
    \label{CC_Corr_Plot}
\end{figure}

The plots in Fig. \ref{CC_Corr_Plot} show our QMC results for the imaginary-time behavior of the current-current correlation function $\widetilde{\Lambda}(\tau_{m})$ defined in Eq. \eqref{cc_corr}. Within the range of the plotted data, the behavior of $\widetilde{\Lambda}(\tau_{m})$ is found to be quite well-defined for both temperature values in the plots. For $\beta=1$, the plot in Fig. \ref{CC_Corr_Plot}(a) shows that $\widetilde{\Lambda}(\tau_{m})\sim b_{0}+b_{2}(\tau-\beta/2)^{2}$ with $b_{0}=\widetilde{\Lambda}(\beta/2)$ and the coefficient $b_{2}$ decreasing as $r$ is increased. As the temperature is lowered, it tends to flatten at $\beta/2$ as shown in the plot (b) for $\beta=4\tsp$, i.e., $\widetilde{\Lambda}(\tau_{m})\sim b_{0}+b_{2}(\tau-\beta/2)^{2}+b_{4}(\tau-\beta/2)^{4}\tsp$. In a narrower imaginary-time scale, $\widetilde{\Lambda}(\tau_{m})$ can become irregular near the central-point at $m=M/2\,$. This is more noticeable for larger $r$ values and at lower temperatures so that, for $\beta\leq 4$ and $r$ tuned closer to the AFM/SDW phase boundary, the numerically discrete current-current correlator behavior with $\tau$ (assuming long imaginary times $\tau\sim\beta/2$) remains reasonably regular. Theoretically, the proxy is expected to approach better the true DC resistivity value as $\beta$ increases\citep{PNAS2017}. However, at lower temperatures $\beta\gtrsim 6$, we found that $r$ needs to be limited even more to ensure that the information for long imaginary times is not hampered by fluctuations that lead to the irregular behavior that we commented on. Hence, in the present study, we chose to focus on the intermediate temperature range $\beta\in[0.25,4]\tsp$.

\begin{figure*}[t]
    \centering
    \includegraphics[width=1.7\columnwidth]{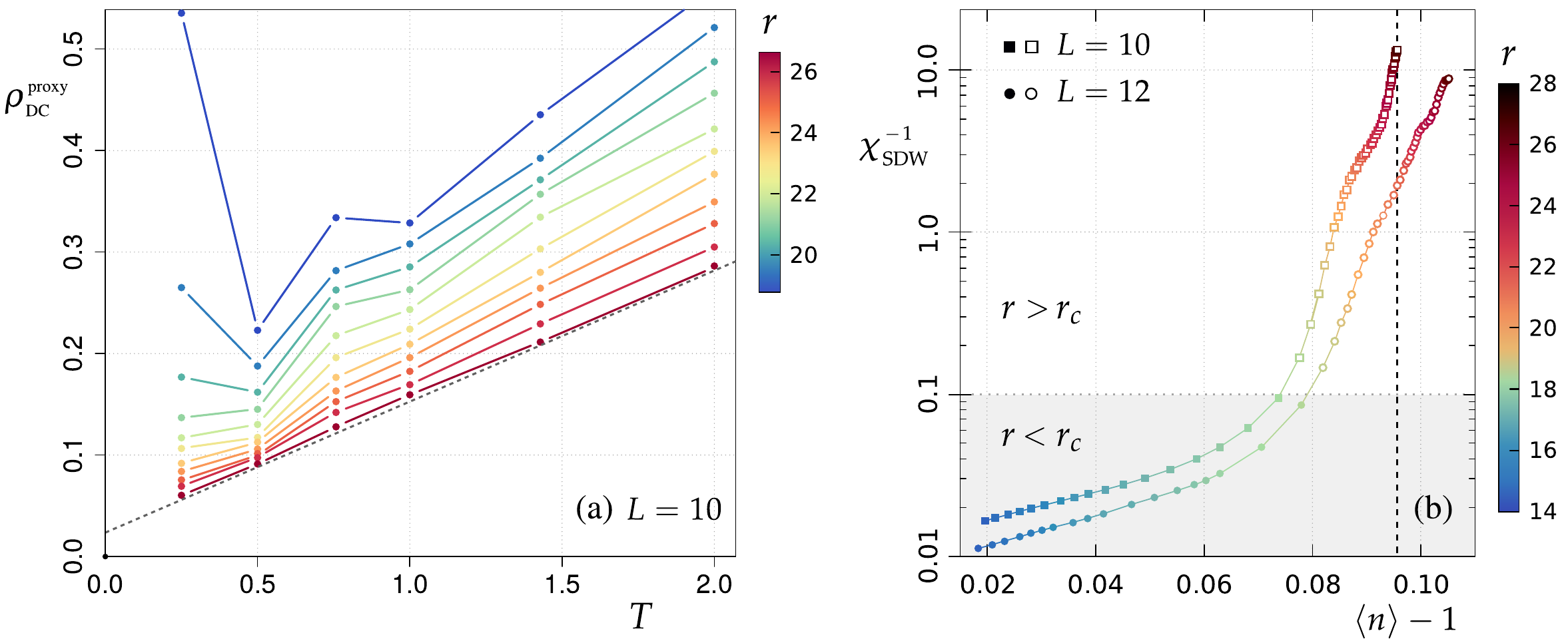}
    \caption{(a) DC resistivity proxy $\rho^{\,\textrm{\tiny proxy}}_{\tsp\textrm{\tiny DC}}$ as a function of $T$ (in units of $\rho_{q}$) for $L=10\tsp$. The $r$ parameter values associated with each set of points in (a) are color-coded according to the vertical bar in the plot. At $r=26.8$, an approximately $T$-linear behavior is obtained. The corresponding fitting function is $\rho^{\,\textrm{\tiny proxy}}_{\tsp\textrm{\tiny DC}}\approx a+b\tsp T\tsp$, where $a=0.025\pm 0.002$ and $b=0.141\pm 0.002\tsp$. (b) The inverse SDW susceptibility $\chi^{-1}_{\tsp\text{\smaller SDW}}$ as a function of the estimated doping $\delta=\langle n \rangle-1$ for $L=10$ and $L=12$ (same fixed temperature: $\beta=4\tsp$). Each point (from both sets) in the plot maps to one $r$ value which is indicated by a color according to the vertical bar on the right-side. {The dashed line marks the optimal doping.}}
    \label{rhodc_results}
\end{figure*}

\begin{figure*}[!t]
    \centering
    \includegraphics[width=2.05\columnwidth]{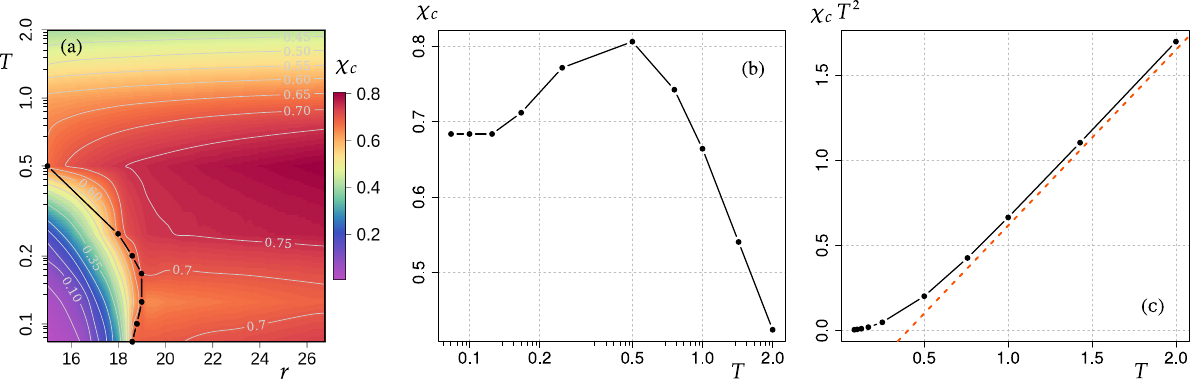}
    \caption{The QMC results obtained for the charge compressibility $\chi_{\tsp c}=\chi^{\text{\smaller\,onsite-CDW}}_{\tsp +}({0})$ are shown in the color-coded plot in (a), where the behavior of this quantity in terms of both $r$ and $T$ (in logarithmic scale) can be visualized in the whole diagram region (the resolution in the $T$-domain was improved via linear interpolation of the QMC data). (b) The dependence of $\chi_{\tsp c}$ for $r=26.8$ as a function of temperature is shown. (c) The dependence of $\chi_{\tsp c}\tsp T^2$ for $r=26.8$ as a function of temperature is shown. (The red dashed line is only a guide for the eye.)}
    \label{q0SUCDW}
\end{figure*}

For all temperature values considered in the present study, we estimated the DC resistivity via the proxy formula given by Eq. \ref{prx2}. In doing so, $[\widetilde{\Lambda}(\beta/2)]^{\tsp 2}$ is taken directly from the QMC results and the second-derivative $\widetilde{\Lambda}^{\prime\prime}(\beta/2)$ is estimated from the fitting function, as explained at the end of Section \ref{sec:III}. In Fig. \ref{rhodc_results}(a), we display the plots for the DC resistivity proxy as a function of the temperature $T$ in the model. In order to better understand how $r$ affects the fermionic system through the Yukawa coupling of the latter with the bosonic order parameter field, we show in Fig. \ref{rhodc_results}(b) a plot of the inverse SDW susceptibility $\chi^{\,\smath{-1}}_{\tsp\text{\smaller SDW}}$ as a function of the average site occupancy $\delta=\langle n \rangle-1\tsp$ defined in the previous section. In the figure, we tune the fermionic system from an AFM/SDW ordered phase into a PM disordered phase by indirectly varying the doping given by $\delta=\langle n \rangle-1\tsp$. {We see that $\delta\gtrsim 0$ when $r<r_{c}$ and also that the SDW susceptibility is strongly suppressed for doping values $\delta\gtrsim 8\%\tsp$.} In our DC resistivity proxy results of Fig. \ref{rhodc_results}(a), an approximately $T$-linear behavior is obtained for a doping of $\delta_{\text{opt}}\approx 0.095$. If we fit the data, we obtain that $\rho^{\,\textrm{\tiny proxy}}_{\tsp\textrm{\tiny DC}}\approx a+b\tsp T\tsp$, where $a=0.025\pm 0.002$ and $b=0.141\pm 0.002\tsp$. Therefore, our results are consistent with the existence of a finite residual resistivity $\rho_0$ at $T=0$. This bears some resemblance with recent transport properties obtained, e.g., in the Hubbard model\footnote{It is worthwhile to point out here that some transport theories of the two-dimensional Hubbard model at weak coupling do not obtain a non-zero intercept at $T=0$ for the resistivity (see, e.g., Refs. \cite{Mueller2021,Ferrero2023}).} in the high temperature regime \cite{Kokalj2017,Kokalj2019}, in other boson-fermion quantum critical theories \cite{Freire2021,Freire2023} and in some Sachdev-Ye-Kitaev-motivated models \cite{Parcollet2020}. Finally, we note that recent experiments also show a finite residual resistivity at $T=0$, e.g., in the cuprate compounds (see Refs. \cite{Cooper2009,Legros2018,Ayres2021}).

The strange metal behavior indicated by our DC resistivity proxy results is established when the doping is close to this latter value, which for our choice of parameters would be the ``optimal doping'' of the model. This means that a strange-metal behavior is indeed obtained for an AFM/SDW quantum critical model at stronger couplings. Moreover, since CO correlations of $s$-wave bond-type are dominant in the PM phase within the temperature range that we ``measured'' the DC resistivity proxy, these fluctuations might also have an influence in the dependence of $\rho_{\tsp\textrm{\tiny DC}}(T)$ at higher temperatures.

The proxy results that we found here support the conclusion that a strange metal phase can emerge from the $Z_2$ spin-fermion model at intermediate temperatures in the critical regime. By contrast, we point out that inside the AFM phase (or reasonably close to it) an upturn of the resistivity is observed. Moreover, we also note that the resistivity of the strange metal phase obtained here does not extend beyond the Mott-Ioffe-Regel limit ($\rho_q\approx 1$) at the measured temperatures. Therefore, we currently see no evidence of the existence of a bad metal regime within the $Z_2$ spin-fermion model.

\subsection{Charge compressibility and charge diffusivity in the strange metal phase}

In the absence of a coupling between the charge and heat carriers, the charge compressibility $\chi_{\tsp c}=\chi^{\text{\smaller\,onsite-CDW}}_{\tsp +}({0})$ (i.e., the $s$-wave onsite-CDW susceptibility for $\bsm{k}={0}\tsp$) and the DC conductivity $\sigma_{\tsp\textrm{\tiny DC}}$ are connected via the Nernst-Einstein relation $\sigma_{\tsp\textrm{\tiny DC}}=D_{c}\,\chi_{\tsp c}$, where $D_{c}$ is the charge diffusion constant\cite{hartnoll2015theory,Kokalj2017,Ferrero2023}. The results of our QMC simulations revealed that $\chi_{\tsp c}$ is weakly dependent on the parameter $r$ when the system is in the disordered/PM phase as shown in Fig. \ref{q0SUCDW}(a). In general, $\chi_{\tsp c}$ is always finite for $r>r_{c}$ (as expected from a metallic system) and, for $r<r_{c}\tsp$, it tends to suppressed as $r$ decreases, inside the AFM/SDW phase. We analyzed this quantity in two regimes: $T\geq 0.75$ (high temperatures) and $T\lesssim 0.5$ (low temperatures). The plots in Figs. \ref{q0SUCDW}(b) and \ref{q0SUCDW}(c) show the overall behavior that we find when $T$ is varied and $r>r_{c}(\beta)$ remains fixed. In the plot of Fig. \ref{q0SUCDW}(b), we see that $\chi_{\tsp c}$ tends to a constant $0.72\pm 0.05$ (this value corresponds to the average of the results for $T\leq 0.5$ in the corresponding figure) for the temperature range $\beta\in[2,12]$. Moreover, in the high temperature regime in Fig. \ref{q0SUCDW}(c), we find that $\chi_{\tsp c}\tsp T^{\tsp 2}$ tends to increase linearly with $T\tsp$, which implies the following fitting function given by $\chi_{\tsp c}\sim a/T+b/T^{\tsp 2}$ with the coefficients being $a\simeq 1.0$ and $b\simeq -0.31\tsp$. For lattice size $L=10$ and $12\tsp$, the overall behavior of the charge compressibility for temperatures $T\geq 0.75$ remains qualitatively the same.

\begin{figure*}[!t]
    \centering
    \includegraphics[width=2.05\columnwidth]{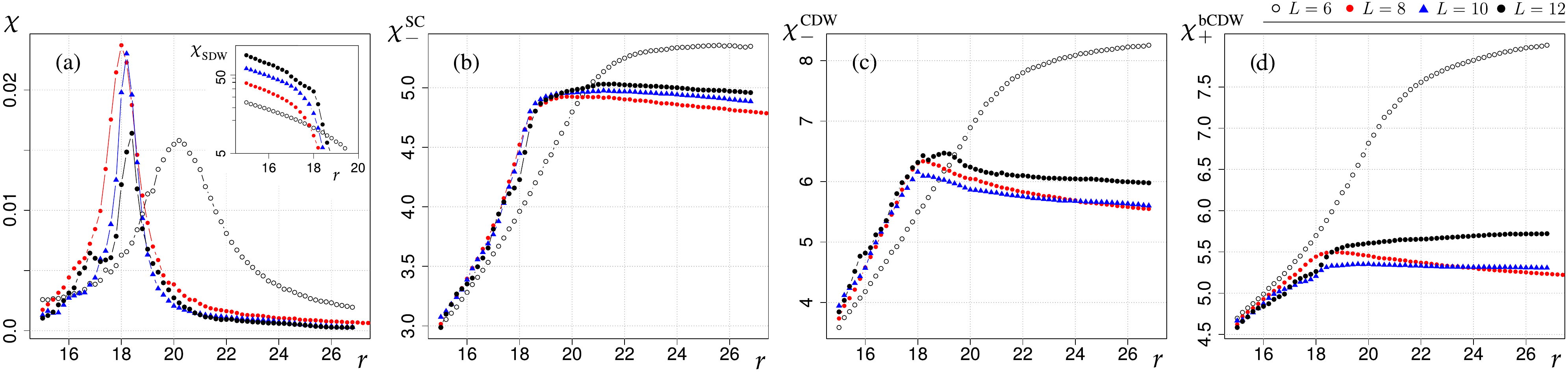}
    \caption{Plots of some susceptibilities estimated by the QMC simulations of the model for $\beta=4$ and lattice size increasing from $L=6$ to $L=12\tsp$: (a) finite-lattice susceptibility $\chi=\chi^{\vspace{1mm}}_{\tsp\text{\smaller SDW}}-\beta N_{s}\tsp\langle\tsp|\mathcal{M}(\varphi)|\,\rangle^{\tsp 2}$ and log-linear plot of the SDW susceptibility $\chi^{\vspace{1mm}}_{\tsp\text{\smaller SDW}}$ (see inset plot), (b) $d$-wave SC susceptibility $\chi_{\tsp -}^{\text{\smaller\,SC}}$, (c) $d$-wave onsite-CDW susceptibility $\chi^{\text{\smaller\,onsite-CDW}}_{\tsp -}(\bsm{k}=\bsm{Q})$, (d) $s$-wave bond-CDW susceptibility $\chi^{\text{\smaller\,bond-CDW}}_{\tsp +}(\bsm{k}=\bsm{0})$.}
    \label{FS_chi}
\end{figure*}

Considering our numerical results from the previous subsection, we showed that an approximately $T$-linear behavior of the proxy for the resistivity $\rho_{\tsp\textrm{\tiny DC}}=1/\sigma_{\tsp\textrm{\tiny DC}}$ extends over a reasonable range of temperatures for the present model. As a result, for lower temperatures, since the charge compressibility tends to a constant value, the charge diffusivity of the model then becomes described by the ``Planckian dissipation''  scaling $D_{c}(T)\sim 1/T$. In this regard, we note that, inspired by groundbreaking results of dissipative processes in holographic models \cite{DT_Son,Hartnoll_Sachdev}, a theory of universal incoherent metallic behavior was proposed in Ref. \cite{hartnoll2015theory}, where it was argued that the transport properties in the strange metal phase that emerges in many strongly-correlated systems should be described in terms of the diffusion of both charge and energy, rather than momentum relaxation. In this latter theory, the mechanism that drives the formation of this non-Fermi liquid 
state at low temperatures is characterized by the charge compressibility saturating to a constant and a charge diffusivity scaling with the inverse of temperature. This scenario is clearly consistent with the scaling that we find in the $Z_2$ spin-fermion model for low temperatures. Therefore, our present result adds support to the interpretation that the mechanism for the approximate $T$-linear resistivity obtained here for the strongly-coupled spin-fermion model at low temperatures might be indeed connected to Planckian dissipation \cite{Zaanen_2019}.

\section{Summary and Outlook}\label{sec:V}

In this work, we have calculated the transport and thermodynamic properties of a two-band spin-fermion model describing itinerant fermions in two dimensions interacting via $Z_2$ antiferromagnetic quantum critical fluctuations by means of a sign-problem-free QMC approach. We have found that this version of the spin-fermion model describes a non-Fermi-liquid metallic regime that exhibits an approximately $T$-linear resistivity above $T_c$ for a strong fermion-boson interaction strength. Using Nernst-Einstein relation, our QMC results has also shown that this strange metal phase is described by either a charge compressibility given approximately by $\chi_{\tsp c}\sim 1/T$ at higher temperatures or by a charge diffusivity consistent with the Planckian dissipation scaling $D_{c}\sim 1/T$ at lower temperatures. We note that both scenarios were recently observed in Ref. \cite{Sci2019} in a study of the two-dimensional Hubbard model on a square lattice for $\text{\emph{U}}=6t$.

It would be interesting to compare our present QMC results with recent more efficient quantum machine learning methods (such as, e.g., Self-learning Quantum Monte Carlo \cite{SLQMC_2017}, Quantum Loop Topography \cite{QLT_2021}, etc) to check if the transport coefficients obtained here are universal for general $O(N)$ spin-fermion models and possibly for other quantum critical theories at strong coupling as well. Moreover, we point out that there are many other directions that can be explored in the future with the current QMC code. For instance, one can further investigate how the prefactor of the $1/T$-dependence of the charge diffusivity at low temperatures calculated in the present work correlates with the choice of other completely different bandstructures in the model, thus potentially establishing the Planckian dissipation as a universal mechanism responsible for the strange metal phase that could emerge in any AFM/SDW quantum critical model at strong coupling. Another direction of research is to calculate the thermal conductivity in the strange metal phase to extract information about the diffusion of heat (via the Nernst-Einstein relation) and to discuss the validity of the Wiedemann-Franz law at low temperatures. Other interesting possibilities include using the current QMC code to study other classes of strongly correlated models such as, e.g., sign-problem-free Hubbard-like models with two bands (for a recent example, see, e.g., Ref. \cite{Christensen_2020}). This investigation could potentially shed light from a numerically exact point of view on the pseudogap phase that emerges in the underdoped regime of the cuprate superconductors.  


\subsection*{Acknowledgments}

R.M.P.T. acknowledges financial support from a CAPES fellowship. The numerical simulations were performed at the SDumont supercomputer of the Laborat\'{o}rio Nacional de Computa\c{c}\~{a}o Cient\'{i}fica (LNCC) in Petrópolis-RJ and at the Laborat\'{o}rio Multiusu\'{a}rio de Computa\c{c}\~{a}o de Alto Desempenho (LaMCAD) of UFG in Goiânia-GO. H.F. acknowledges funding from CNPq under Grant No. 311428/2021-5. 

\appendix

\section{Finite-size effects}\label{AppendixA}

In the main text, we pointed out that the QMC simulations of the metallic system described by the spin-fermion model depend on a finite size $L$ of the lattice. In this appendix, we will present an analysis for different lattice sizes in order to show that the finite-size effects are indeed mild for all the quantities calculated in this work.

\begin{figure}[!t]
    \centering
    \includegraphics[width=0.8\columnwidth]{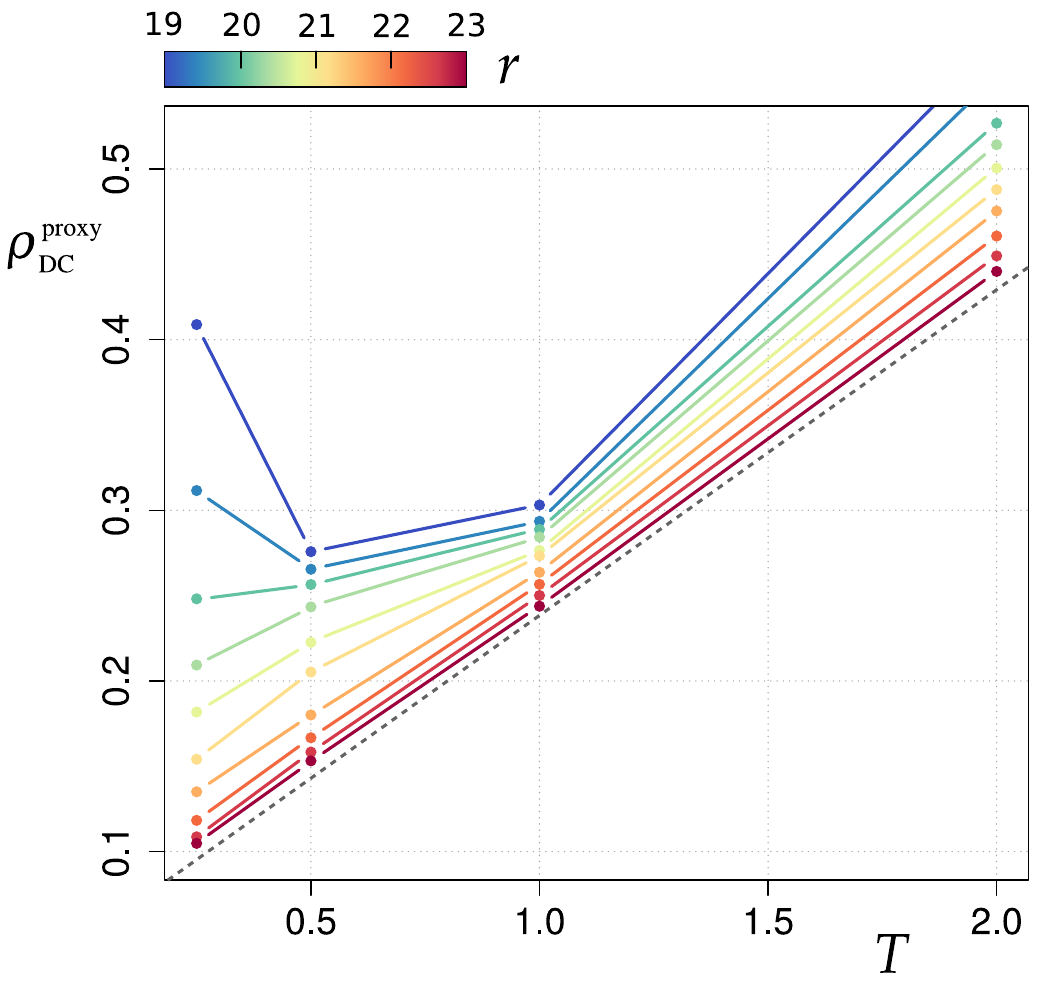}
    \caption{DC resistivity proxy $\rho^{\,\textrm{\tiny proxy}}_{\tsp\textrm{\tiny DC}}$ as a function of $T$ (in units of $\rho_{q}$) for $L=12\tsp$. The $r$ values associated with each $T$-plot are color-coded according to the horizontal bar in the plot. The corresponding fitting function is $\rho^{\,\textrm{\tiny proxy}}_{\tsp\textrm{\tiny DC}}\approx a'+b'\tsp T$, where $a'=0.074\pm 0.002$ and $b'=0.166\pm 0.002$.}
    \label{rhoDC_L12}
\end{figure}

Let us start by considering the finite-lattice susceptibility $\chi$ and the SDW susceptibility $\chi^{\vspace{1mm}}_{\tsp\text{\smaller SDW}}$ at $\beta=4\tsp$. In Fig. \ref{FS_chi}(a), we see that the peak of $\chi$ is broad for a small lattice size ($L=6\tsp$). Therefore, it leads to an estimate for $r_{c}$ that deviates from the more precise estimates obtained for larger lattices. Indeed, for $L\geq 8\tsp$, the peak becomes much sharper and only shifts slightly to the right as we increase $L\tsp$. Also, we notice that the maximum value of $\chi$ tends to decrease, although the logarithmic of $\chi^{\vspace{1mm}}_{\tsp\text{\smaller SDW}}$ increases for $r<r_{c}$ (see the inset plot in the same figure). Thus, these results indicate that the upper part of the magnetic phase diagram from Fig. \ref{L8_diagram}(a) does not change much for larger systems. After analyzing the behavior of the two main fluctuations competing in the PM phase, namely, the $d$-wave SC and the CO fluctuations of $s$-wave bond-CDW-type, we find that these quantities are also only weakly affected by the lattice size for $r<r_{c}$ and $L\geq 8$, as shown in the plots of Figs. \ref{FS_chi}(b)-(d). From these plots, we see that $\chi_{\tsp -}^{\text{\smaller\,SC}}$ tends to scale linearly with $L$, as $r$ is tuned away from $r_{c}$ in PM phase. The same does not apply though to the CO susceptibilities $\chi^{\text{\smaller\,CDW}}_{\tsp -}(\bsm{k}=\bsm{Q})$ and $\chi^{\text{\smaller\,bCDW}}_{\tsp +}(\bsm{k}={0})$, shown in Figs. \ref{FS_chi}(c) and (d), respectively. However, for moderate temperatures, the results for lattice sizes $L\geq 10$ presented here indicate that the latter two quantities might increase with $L$ at a faster rate when compared with $\chi_{\tsp -}^{\text{\smaller\,SC}}$ in the PM phase, such that the regime with dominant CO correlations at moderate temperatures described in the main text will also likely remain in the thermodynamic limit.

For completeness, regarding the QMC results for the DC resistivity proxy of the spin-fermion model, we consider the plot in Fig. \ref{rhoDC_L12} for lattice size $L=12\tsp$. Compared to Fig. \ref{rhodc_results}(a) in the main text, a similar trend is observed in this plot with a regime inside the AFM phase (or reasonably close to it) displaying an upturn of the resistivity as a function of $T$. Upon doping, a quantum critical regime with an approximately $T$-linear resistivity emerges in the model, which is found for a doping parameter reasonably close to the optimal value $\delta_{\text{opt}}$ obtained for $L=10$ (see also Fig. \ref{rhodc_results}(b)). In this regime, we obtain a linear fitting function given by $\rho^{\,\textrm{\tiny proxy}}_{\tsp\textrm{\tiny DC}}\approx a'+b'\tsp T$, where $a'=0.075\pm 0.002$ and $b'=0.166\pm 0.002\tsp$. Therefore, we conclude that the finite-size effects for the DC resistivity proxy are also relatively mild for the estimate of this transport coefficient in the present model.

\end{document}